\documentclass[a4paper,11pt]{article}
\usepackage{jcappub}
\usepackage{aas_macros}
\usepackage{orcidlink}
\usepackage{bm}
\usepackage{mathrsfs}
\usepackage{subcaption}

\def\beq{\begin{equation}}
\def\eeq{\end{equation}}
\def\be{\begin{equation}}
\def\ee{\end{equation}}
\def\bea{\begin{eqnarray}}
\def\eea{\end{eqnarray}}

\renewcommand\k{{\bf k}}
\newcommand\q{{\bf q}}

\newcommand{\khat}{\hat{k}}
\newcommand{\qhat}{\hat{q}}

\newcommand{\accel}{ACCEL$^{2}$}
\newcommand{\lya}{Ly$\alpha$}
\newcommand{\poned}{P_{\mathrm{1D}}}
\newcommand{\pthreed}{P_{\mathrm{3D}}}
\newcommand{\plin}{P_\mathrm{lin}}
\newcommand{\pthreedeft}{P_{\mathrm{3D}}^\mathrm{EFT}}
\newcommand{\skm}{s~km$^{-1}$}
\newcommand{\kms}{km~s$^{-1}$}
\newcommand{\hmpc}{$h~$Mpc$^{-1}$}
\newcommand{\mpch}{$h^{-1}~$Mpc}

\newcommand{\tp}{\mathrm{T}}
\newcommand{\lnAs}{\ln (10^{10} A_s)}

\defcitealias{collaborationPlanck2018Results2020}{Planck2018}

\makeatletter

\gdef\@fpheader{Prepared for submission to JCAP \hfill MIT-CTP/6027}
\makeatother

\title{\boldmath Analytic compression of the effective field theory of the Lyman-alpha forest}

\author[1,2,3]{{Naim~G\" oksel Kara\c{c}ayl{\i}}\orcidlink{0000-0001-7336-8912},}
\author[4,5]{{Mikhail M. Ivanov}\orcidlink{0000-0002-6745-984X},}
\author[4,5]{{Roger de Belsunce}\orcidlink{0000-0003-3660-4028},}
\author[6]{{Corentin Ravoux}\orcidlink{0000-0002-3500-6635},}
\author[7]{{Jean M. Sexton}\orcidlink{0000-0003-2551-1678},}
\author[7]{{Zarija Luki\' c}}

\affiliation[1]{Center for Cosmology and AstroParticle Physics, The Ohio State University, 191 West Woodruff Avenue, Columbus, OH 43210, USA}
\affiliation[2]{Department of Astronomy, The Ohio State University, 4055 McPherson Laboratory, 140 W 18th Avenue, Columbus, OH 43210, USA}
\affiliation[3]{Department of Physics, The Ohio State University, 191 West Woodruff Avenue, Columbus, OH 43210, USA}
\affiliation[4]{Center for Theoretical Physics -- a Leinweber Institute, Massachusetts Institute of Technology, Cambridge, MA 02139, USA}
\affiliation[5]{The NSF AI Institute for Artificial Intelligence and Fundamental Interactions, Cambridge, MA 02139, USA}
\affiliation[6]{Universit\'{e} Clermont-Auvergne, CNRS, LPCA, 63000 Clermont-Ferrand, France}
\affiliation[7]{Lawrence Berkeley National Laboratory, 1 Cyclotron Rd, Berkeley, CA 94720, USA}
\emailAdd{karacayli.1@osu.edu}

\abstract{
The 1D flux power spectrum ($P_{\mathrm{1D}}$) of the Ly$\alpha$ forest provides an exceptionally high-resolution probe of structure formation down to small scales ($k\approx1-10~\text{$h~$Mpc$^{-1}$}$). These scales carry the imprints of massive neutrinos, warm dark matter, and the running of the primordial power spectrum spectral index. The effective field theory (EFT) is a promising perturbative approach to systematically and efficiently describe the Ly$\alpha$ forest, but it faces challenges in its application to $P_{\mathrm{1D}}$, as many EFT parameters become degenerate when projected along the line of sight. In addition, this projection generates new stochastic terms from the integration over small-scale modes. In this work, we address these issues by compressing the EFT model space using the Fisher matrix formalism and linearizing the resulting compression directions, enabling analytic template marginalization and significantly reducing the computational cost of likelihood evaluation. We use hydrodynamical simulations to obtain a baseline estimate of EFT parameters, and use the DESI DR1 $P_{\mathrm{1D}}$ measurements to derive compression directions. We then marginalize over deviations from the baseline using these compression directions and forecast the constraining power of our formalism. We find that even in conservative scenarios where each data redshift bin requires its own set of EFT parameters, the cosmological constraints saturate with the linear bias, two leading-order 1D stochastic terms, and three principal combinations of the remaining EFT templates.
In this case, our forecasted precision of the amplitude ($\Delta^2_p$) and the logarithmic slope ($n_p$) of the linear matter power spectrum at the pivot scale ($k_p=0.7~\text{Mpc}^{-1}$) is $10\%$ and $2.0\%$, respectively, which is similar to emulator-based analyses that include observational data systematics.
}

\begin{document}
\maketitle
\flushbottom

\section{Introduction\label{sec:intro}}
The standard model of cosmology has been remarkably successful at explaining high-precision measurements of the cosmic microwave background (CMB) anisotropies \cite{collaborationPlanck2018Results2020}. Together with the distance measurements from baryon acoustic oscillations (BAO) and supernovae, the standard model does an excellent job in constraining the expansion history and the large-scale structure of the universe \cite{desiY3LyaBAO2025, desiY3BaoAndCosmology2025, pantheonSupernovae2022, unionSupernovae2025, desY5Supernovae2024}. While tensions remain within these datasets, such as the Hubble tension, which could indicate new physics, and while a new (time-evolving) picture of the nature of dark energy unassertively emerges \cite{desiY3BaoAndCosmology2025}, these datasets leave small-scale ($k \gtrsim 0.2~\text{\hmpc}$) matter clustering underexplored. These are the scales where the imprints of massive neutrinos \cite{croftCosmologicalLimitsNeutrino1999, palanqueDelabrouilleNeutrinoMass2015, yecheNeutrinoMassesXQ2017}, warm dark matter \cite{narayananWDMLyaForest2000, seljakSterileNeutrinosDM2006, wangLyaDecayingDM2013, irsicFuzzyDMfromLya2017, boyarskyLyaWDM2009, vielWarmDarkMatter2013, baurLyaCoolWDM2016, irsicConstraintsWDM2017, villasenorWarmDarkMatter2023}, and a running spectral index \cite{vielPrimordialPowerSpectrumLya2004} are expected to be most discernible through structure formation.

Most recently, these scales are measured with exquisite precision in the 1D flux power spectrum ($\poned$) of the Lyman-$\alpha$ (\lya) forest between $2.2 \leq z \leq 4.4$ from the Dark Energy Spectroscopic Instrument (DESI) data release (DR) 1 \cite{karacayliQmleP1dDesiDr12024, ravouxFFTP1dDesiDr12024}. This measurement then improved the upper bound on the sum of neutrino masses by $20\%$ relative to CMB-alone and improved the precision on the running of the spectral index by $50\%$ \cite{desiP1dInferenceDr1_2026}. 

The \lya\ forest is a series of absorption lines in quasar spectra caused by resonant scattering with intervening neutral hydrogen. At these redshifts, the Universe is relatively young, such that each absorption line traces mildly nonlinear fluctuations in the matter density field. However, a rigorous theoretical interpretation of the Ly$\alpha$ $\poned$ remains nontrivial, because of its sensitivity to the thermal and ionization state of the intergalactic medium (IGM). Conventional $\poned$ inferences rely on computationally expensive hydrodynamic simulations,
limiting the parameter space that can be reliably explored \cite{chabanierOnedimensionalPowerSpectrum2019, fernandezCosmologicalConstraintsEboss2024, desiP1dInferenceDr1_2026}. 

Recently, the effective field theory (EFT;~\cite{McDonald:2009dh, Baumann:2010tm, Carrasco:2013mua}) of large-scale structure has emerged as a systematic and computationally efficient framework for analytically describing the large-scale structure on mildly nonlinear scales. EFT parameters are a set of bias and counterterm parameters that are bounded by the symmetries relevant to the analyzed tracer \cite{ivanovEffectiveLya2024, deBelsunce:2025edy}.
When restricted to the one-loop order, EFT is shown to accurately describe the 3D \lya\ forest power spectrum ($\pthreed$) up to scales $k_\mathrm{max}\approx 3~\text{\hmpc}$~\cite{ivanovEffectiveLya2024, belsuncePrecisionEft2025, Hadzhiyska:2025cvk}. Consequently, EFT of $\poned$ faces a difficult challenge since the calculation of $\poned$ formally involves integration over UV modes beyond the cut-off scale:
\begin{equation}
    \poned (k_\|) = \int_{k_\|}^\infty \frac{q \, \mathrm{d}q}{2\pi} \pthreed(q, k_\|)\, ,
\end{equation}
where $k_\parallel$ is the wavenumber along the line of sight. A solution is to perform the integration up to $k_\mathrm{max}$ where the EFT description works, and to absorb the UV sensitivity into 1D stochastic terms, i.e. $\poned(k_\|) = \poned^\mathrm{EFT} + \poned^\mathrm{stoc}$. The functional form of $\poned^\mathrm{stoc}(k_\|)$ is fixed by symmetries to be a polynomial of $k_\|^2$. At lowest orders, it is well approximated as a fourth-order polynomial. The coefficients of this polynomial should be either matched to data or simulations. Fitting this polynomial to the data may remove a substantial part of the shape information available in $\poned(k_\|)$.

This point ties into the next major issue with EFT of $\poned$---its high degree of freedom, with 18 bias parameters, $b_\mathcal{O}$, and a highly restricted available momentum range. One would expect to fit a wide variety of data vectors without even modifying the fundamental cosmological parameters with such a large parameter space. In particular, for 1D statistics, where the anisotropic signal is integrated over, the EFT parameters become highly degenerate and unconstrained. This can be partially addressed by priors derived from simulations. However, another complication that follows is the joint analysis of multiple redshift bins, which would require 18 parameters for each bin. Without assumptions of time-dependence of EFT parameters, this would force one to work in a model space with 180 free parameters for a $\poned$ measurement between $2.2\leq z\leq 4.0$ in ten bins.

Even though the EFT of $\poned$ faces these problems, it remains highly flexible, allowing efficient exploration of within- and beyond-$\Lambda$CDM models, including neutrino self-interactions \cite{heFreshNeutrinoSelfInteractions2025} and sterile neutrino production mechanisms \cite{parashariLyaForestBoundsSterileNeutrino2026}, and the abundance of primordial black hole as dark matter \cite{ivanovEftPrimordialBlackHoles2026}. In this work, we push the EFT to its limit at ever smaller scales to be applicable to $\poned$ analyses. Our proposed workflow has three steps. (1) We employ the EFT parameter space relations of $b_\mathcal{O}=b_\mathcal{O}(b_1)$, which leaves only one free parameter of the deterministic EFT model~$b_1$, which is empirically derived from simulations~\cite{ivanovEftEboss2024, belsuncePrecisionEft2025}. (2) We relax this model by identifying the most important modes of deviations from the simulation-based $b_\mathcal{O}(b_1)$ relation.  
We compress the model space by selecting a few eigenvectors with the largest eigenvalues using a Fisher matrix formalism. The compressed directions, which we denote $\bm q_n$, correspond to combinations of $b_\mathcal{O}$ that affect $\poned$ most significantly, i.e., which are most detectable given the data covariance matrix. (3) In this Fisher analysis, we explicitly linearize the EFT model of $\poned$ with respect to $\bm q_n$ around the fiducial point. This enables us to apply analytic template marginalization, removing all $\bm q_n$ from the ``free" parameter space.

Our work draws from other works on analytic methods for likelihood evaluation that address the ``many parameters" problem in inference \cite{taylorAnalyticMethodsLikelihood2010, hadzhiyskaEfficientMarginalisation2023}. These works build general-purpose likelihoods around the maximum (or a fiducial) point. Such a scheme would also solve the many-parameter problem of EFT of $\poned$ in terms of computational efficiency. However, one would be left to investigate the appropriate choice of priors and which parameters most influence the analysis among 18 parameters per redshift bin. Our model compression addresses this problem before the inference begins. The compressed vectors are readily sorted by importance and are orthogonal by construction. So, incremental inclusion of compressed directions has a well-behaved, likely convergent, impact on the final analysis results. Parameters not removed after compression have the weakest influence on the analysis and are effectively fixed to fiducial values.

In many works on large-scale structure analysis, compression typically addresses covariance estimation for large data vectors \cite{heavensMassiveLosslessDataCompression2000, gualdiEnhancingBossBispectrum2019, philcoxFewerMocksLessNoise2021}. By reducing the dimensions of the data vector, one requires fewer independent samples to construct a less noisy covariance matrix. That is not, so far, a problem for DESI's $\poned$. Its relatively simple configuration allows for data-driven methods to be sufficient for robust covariance matrix estimation. For example, bootstrap sampling of 300,000 quasars in DESI DR1 yields an accurate covariance matrix with well-motivated smoothing, which is validated using many mock realizations \cite{karacayliDesiY1P1dValidation}. To make our formalism clear, our core framework compresses the \textit{model} parameter space rather than the data vector space.

This paper is organized as follows. In Section~\ref{sec:data}, we describe the hydrodynamical simulations used to calibrate EFT parameters, and DESI DR1 $\poned$ measurement used to derive compression directions and forecast the constraining power of our formalism.
Section~\ref{sec:eftmodel} is the core section where we formulate our compression scheme after overviewing EFT of $\poned$ and deriving $b_\mathcal{O}$ and relations $b_\mathcal{O}(b_1)$. 
We illustrate how our formalism can be implemented in Section~\ref{sec:likeli}. This section provides an outline of analytic template marginalization and highlights important approximations that significantly reduce the computational cost of the EFT of $\poned$.
We show how well cosmological parameters can be constrained by the incremental addition of nuisance parameters, using the DESI DR1 measurements' covariance matrix in a forecast setting in Section~\ref{sec:forecast}. 
Lastly, we discuss the limitations of our formalism and directions for future work in Section~\ref{sec:discuss}.

\section{Data\label{sec:data}}
Our work relies on simulation data to calibrate EFT parameters, and on real $\poned$ data to derive compression directions and assess the predictive power of our formalism. In this section, we describe the simulated and real data (DESI DR1 $\poned$ measurements) used in our work.

\subsection{Simulations\label{sec:simulation}}
We fit the EFT parameters to simulations to investigate relations as a function of $b_\mathcal{O}=b_\mathcal{O}(b_1)$. The \accel\ simulations form the basis for this study, while Sherwood simulations serve as a variation and a cross-check. We provide an overview of \accel\ and Sherwood simulations below.

\accel\ simulations are generated using the \texttt{Nyx} code, which is a highly-parallel, adaptive mesh hydrodynamics solver for cosmological simulations \cite{almgren2013Nyx, Sexton2021Nyx}. We operate on the highest-resolution box of \accel\ simulations to ensure $\poned$ convergence \cite{chabanierAccel2Simulations2024}.  There are $6144^3$ hydrodynamical elements and dark matter particles in this box, with a side length of $L=160~$\mpch, yielding an effective resolution of $25~h^{-1}~$kpc. The cosmological parameters are $\Omega_m=0.31, \Omega_b=0.0487, h=0.675, n_s=0.96,$ and $\sigma_8=0.83$ based on \emph{Planck} 2015 results \cite{planck2016CosmologyParameters}. We use five snapshots at $z=2.0, 2.6, 3.0, 3.6, 4.0$. The $\pthreed$ is measured using FFT in four $\mu$ bins and with bin width $\Delta k=0.04~\text{\mpch}$ using the \texttt{gimlet} software \cite{gimlet2016}. This is the true, albeit $k$- and $\mu$-averaged, $\pthreed$ since all Fourier modes within the box are included.

The Sherwood simulations are high-resolution hydrodynamic simulations generated using the smoothed particle hydrodynamics code \texttt{P-Gadget3} \cite{boltonSherwoodSimulationSuite2017}, which is a modified version of \texttt{Gadget-2} \cite{springelGadget2_2005}. We use the \lya\ forest extractions of ref.~\cite{givansNonlinearitiesCross2022}. As noted in ref.~\cite{ivanovEftEboss2024}, these simulations do not match observed $\poned$; therefore, the UVB radiation amplitude needs to be scaled for a better match in the optical depth and $\poned$.
There are $2\times 1024^3$ particles in the box in a side length of $L=160~$\mpch. As these numbers indicate, \accel\ has an improved physical resolution by a factor of 6 in the same cosmological volume \cite{chabanierAccel2Simulations2024}. As refs.~\cite{chabanierAccel2Simulations2024, lukicLymanAlphaForestHydroSims2015} have shown, this resolution is not sufficient to achieve sub-percent convergence for the 1D and 3D power spectra, which is the main reason why the \accel\ simulation suite is our baseline. The input cosmology for Sherwood is based on \emph{Planck} 2013 results \cite{Planck2013CosmologicalParameters}: $\Omega_m=0.308, \Omega_b=0.0482, h=0.678, n_s=0.961,$ and $\sigma_8=0.829$, which is only a minor deviation from the \accel\ input cosmology.

For both simulations, the $\pthreed$ is averaged over three directions and assigned a diagonal covariance matrix based on Gaussian errors calculated from the total number of Fourier modes ($N_i$) in each bin $i$. To mitigate excess weight on smaller scales, we introduce an ad hoc  5\% noise floor for simulation uncertainties, as per refs.~\cite{mcdonaldTowardMeasurementCosmological2003, chabanierAccel2Simulations2024}: $w_i^{-1/2}=P_i (\sqrt{2 /N_i} + 0.05)$.\footnote{Without a noise floor, the model fits would effectively ignore low $k$ points.} When obtaining the stochastic terms, however, we treat each $\poned$ mode equally. Even though we have access to modes up to the Nyquist scale in simulations, EFT is a perturbation theory whose one-loop computation is valid up to an effective scale, $k_\mathrm{max}$. Therefore, we limit the maximum wavenumber to $k_\mathrm{max}=3~$\hmpc\ for all redshift bins for both simulations.

\subsection{DESI measurements}
DESI observed over 1.5 million quasars in its first year of operations \cite{desiKp2DataRelease12024, ashleyDesiLssCatalog2024}. Of all these quasars, 450,000 are at $z > 2.1$, so that the \lya\ forest region falls within DESI's wavelength coverage of 3600--9800~\AA. Using this immense sample, ref.~\cite{karacayliQmleP1dDesiDr12024} measured $\poned$ using the optimal estimator from $z=2.2$ to $z=4.4$ in 12 bins, while in a companion paper ref.~\cite{ravouxFFTP1dDesiDr12024} applied the Fast Fourier Transform approach. The underlying estimator differentiates these two measurements. Because the optimal estimator is robust against the major systematics of the \lya\ forest, specifically masking and continuum fitting, we use its results from the high-SNR sample, where the average per pixel $\mathrm{SNR} > 3$ in the \lya\ forest region is applied to quasar selection. This high-SNR sample contains 62,807 quasars, which increases statistical errors but nearly eliminates noise-induced systematics, and is expected to have higher purity and completeness for identified astrophysical contaminants.

Ref.~\cite{karacayliQmleP1dDesiDr12024} measures the DESI DR1 $\poned$ in 80 $k$ bins in each of the 12 redshift bins. The reach of these $k$ bins is limited by DESI's spectrograph resolution and continuum fitting errors. Additionally, $z=4.2$ and $z=4.4$ bins from the $\mathrm{SNR} > 3$ sample have unreliable error estimates due to low statistics. The recommended redshift $(2.2\leq z \leq 4.0)$ and scale ($10^{-3}~\text{\skm}<k<0.5\pi/R_z$, where $R_z\equiv c \Delta \lambda_\mathrm{DESI}/(1+z)\lambda_\text{\lya}$ and $\Delta\lambda_\mathrm{DESI}=0.8~$\AA) cuts leaves 609 data points in total. However, EFT is valid in scales up to $3~\text{\hmpc}$, and therefore, as we note in section~\ref{sec:forecast}, we use a more conservative scale cut, leaving us with 37 $k$ bins per redshift.

\section{Effective field theory of \texorpdfstring{$\poned$}{P1D} \label{sec:eftmodel}}
The complete 18-parameter space for $\poned$ consists of 12 parameters at the one-loop order, 3 counter terms, and 3 stochastic terms.
The EFT computation of the deterministic part of the 3D \lya\ power spectrum at redshift $z$ at the one-loop order can be summarized as: 
\be 
\pthreedeft\left(k, \mu\right)=K^2_1(\mu)P_{\rm lin}(k) + P_{22}(k, \mu) + 2P_{13}(k, \mu) 
+P_\mathrm{ctr}^\mathrm{EFT}(k, \mu)
\,,
\ee 
where $K_1(\mu)\equiv (b_1-b_\eta f\mu^2)$, $f=\frac{d\ln D_+}{d\ln a}$ is the logarithmic growth factor, $b_1(z),b_\eta(z)$ are linear bias parameters. The explicit expressions for the one-loop terms $P_{22}$ and $P_{13}$ are presented in Appendix~\ref{app:EFT_kern} and the original papers~\cite{ivanovEftEboss2024,ivanovEffectiveLya2024} (see  also~\cite{Desjacques:2018pfv}).

The one-loop EFT perturbative expansion has 12 bias parameters (formally 13, but we drop one as discussed below):
\begin{itemize}
    \item Two parameters at linear order: $b_1$ and $b_\eta$,
    \item Six at quadratic order: $b_2, b_{\mathcal{G}_2}, b_{(KK)_\|}, b_{\Pi^{[2]}_\|}, b_{\delta\eta},$ and $b_{\eta^2}$,
    \item Four at cubic order: $b_{\Pi^{[3]}_\|}, b_{(K \Pi^{[2]})_\|}, b_{\delta \Pi^{[2]}_\|},$ and $b_{\eta \Pi^{[2]}_\|}$.\footnote{At this order, the $b_{\Gamma_3}$ term is formally present, but, in what follows, we set $b_{\Gamma_3}=0$ due to the degeneracy with $b_{\mathcal{G}_2}$ \cite{ivanovEffectiveLya2024}. This could be alleviated by including cross-correlation with matter, halos, or quasars (see, e.g.,~\cite{Chudaykin:2025gsh}). We leave this to future work.}
\end{itemize}
In addition, there are three higher derivative counterterms, which we parameterize as 
\begin{equation}
    P_\mathrm{ctr}^\mathrm{EFT}=-2(c_0 + c_2\mu^2 + c_4\mu^4) K_1(\k) (k^2\plin/A_\mathrm{ctr})\,,
\end{equation}
where $c_n$ are the free counterterm parameters, and $A_\mathrm{ctr}=10~\text{\mpch}$ is a dimensionality constant so that $c_n\sim b_1$. 

$\poned$ is further complicated by projection integration up to formally infinitely small scales, $k\rightarrow\infty$. Practically, the integration is performed up to the EFT scale-cut, $k_\mathrm{max}$, and contributions above this scale are renormalized by the stochastic counterterms that are described by a polynomial:
\begin{equation}
    \label{eq:master_eft_p1d}\poned(k_\|) = \poned^\mathrm{EFT} + \poned^\mathrm{stoc} = \int_{k_\|}^{k_\mathrm{max}} \mathrm{d}q \frac{q \pthreedeft(q, \mu=k_\|/q)}{2\pi} + \sum_{n=0}^{n_\mathrm{max}} \mathcal{C}_{2n} (k_\|/k_\mathrm{stoc})^{2n},
\end{equation}
where $\mathcal{C}_n$ are the stochastic terms and we set $k_\mathrm{stoc}=5~\text{\hmpc}$. This is the main EFT equation. In Section~\ref{sec:biasrelations}, we will find three $\mathcal{C}_n$, namely $\mathcal{C}_0$, $\mathcal{C}_2$, and $\mathcal{C}_4$, are sufficient to quantify $\poned^\mathrm{stoc}$ using simulations.

Our formulation slightly differs from ref.~\cite{belsuncePrecisionEft2025} as we absorb the 3D shot noise ($P_\mathrm{shot}$) and scale-dependent stochasticity Wilson coefficients ($a_{0,2}$) into $\mathcal{C}_n$ because the contributions of terms into $\poned$ are  100\% degenerate with the 1D stochastic contributions~\cite{ivanovEffectiveLya2024}.

Eq.~\eqref{eq:master_eft_p1d} evaluates $\poned$ in \mpch\ units, whereas the observed $\poned$ is in velocity units. The conversion factor between the two depends on redshift and cosmology, and is as follows:
\begin{equation}
    \label{eq:mpc2kms}g(z) =  100 \frac{E(z)}{1+z} ~\frac{\text{\kms}}{\text{\mpch}} = 100 \frac{\sqrt{\Omega_\Lambda + \Omega_m (1+z)^3}}{1 + z} \frac{\text{\kms}}{\text{\mpch}},
\end{equation}
where we assume zero curvature such that $\Omega_\Lambda=1-\Omega_m$. Then, $\poned^\text{\kms}=g(z)\poned^\text{\mpch}$ and $k_\text{\mpch} = g(z) k_\text{\kms}$. Through this conversion, $\Omega_m$ affects both the amplitude and the shape of $\poned$.

\subsection{Bias relations\label{sec:biasrelations}}
To first order, EFT bias parameters follow a simple linear relation as a function of $b_1$: $b_\mathcal{O}=b_\mathcal{O}(b_1)$ \cite{ivanovEffectiveLya2024, belsuncePrecisionEft2025}. Using the simulations described in Section~\ref{sec:simulation}, we first find the numerical values for these bias parameters $b_\mathcal{O}$, and then investigate the relationship $b_\mathcal{O}=b_\mathcal{O}(b_1)$ in this section.

We directly fit all 15 bias parameters of $\pthreed$\footnote{Note that three stochastic terms from Eq.~\eqref{eq:master_eft_p1d} are for the $\poned$ only.} using \texttt{iminuit} \cite{iminuit}. Our preliminary fits yielded valid solutions, but indicated numerical difficulties in the computation of the Hesse (covariance) matrix. These numerical difficulties arise from near-degeneracies among parameters and the large number of free parameters; they are mitigated as follows. We first introduce Gaussian priors of $\mathcal{N}(0, 4)$ on all ten one-loop bias parameters and tighter Gaussian priors of $\mathcal{N}(0, 1)$ on the three counterterms \cite{belsuncePrecisionEft2025}, while not imposing any priors on the linear order bias parameters $b_1$ and $b_\eta$ as listed in Table~\ref{tab:bias_sim_priors}. We confirmed that the best-fitting values are consistent with these priors. To address the large parameter space, we use the minimization strategy that approximates the Hesse matrix, following the \texttt{iminuit} guideline.\footnote{This is the \texttt{strategy=0} setting for readers who are familiar with \texttt{iminuit}.}

\begin{table}
    \centering
    \begin{tabular}{|c|c|}
        \hline
         Parameter & Prior \\
        \hline
        $b_1, b_\eta$ & None \\
        $b_2, b_{\mathcal{G}_2}, b_{(KK)_\|}, b_{\Pi^{[2]}_\|}, b_{\delta\eta}, b_{\eta^2}, b_{\Pi^{[3]}_\|}, b_{(K \Pi^{[2]})_\|}, b_{\delta \Pi^{[2]}_\|}, b_{\eta \Pi^{[2]}_\|}$ & $\mathcal{N}(0, 4)$ \\
        $c_0, c_2, c_4$ & $\mathcal{N}(0, 1)$ \\
        \hline
    \end{tabular}
    \caption{Priors for the EFT bias parameters to stabilize the numerical issues in the Hesse matrix computation. These are from ref.~\cite{belsuncePrecisionEft2025}, except in our definition $c_n$ are scaled up by a factor of $A_\mathrm{ctr}=10~\text{\mpch}$.}
    \label{tab:bias_sim_priors}
\end{table}

These numerical challenges and approximations then require us to apply a Monte Carlo technique to quantify the error on the best-fitting parameters. To achieve higher accuracy in our error estimation, we perform 200 Monte Carlo fits and jitter the initial minimization point to explore a broader chi-squared profile for each fit. One could also perform fits directly at the field level to reduce uncertainty in the (quadratic) bias parameters \cite{deBelsunce:2025bqc}, benefiting from cosmic variance cancellation -- we leave this to future work.

\begin{figure}
    \centering
    \includegraphics[width=\linewidth]{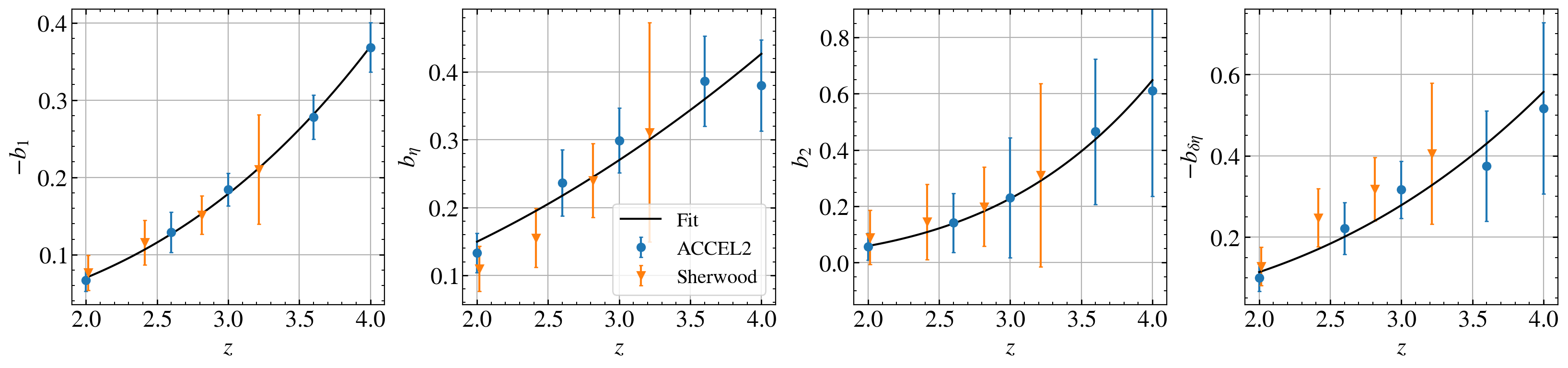}

    \caption{Four EFT bias parameters measured from \accel\ ({\it blue circles}) and Sherwood ({\it orange triangles}) simulations at different redshifts. Both simulations agree within the error bars. A power law provides a good description for the redshift evolution ({\it black line}).
    }
    \label{fig:bias_vs_z}
\end{figure}

\begin{figure}
    \centering
    \includegraphics[width=\linewidth]{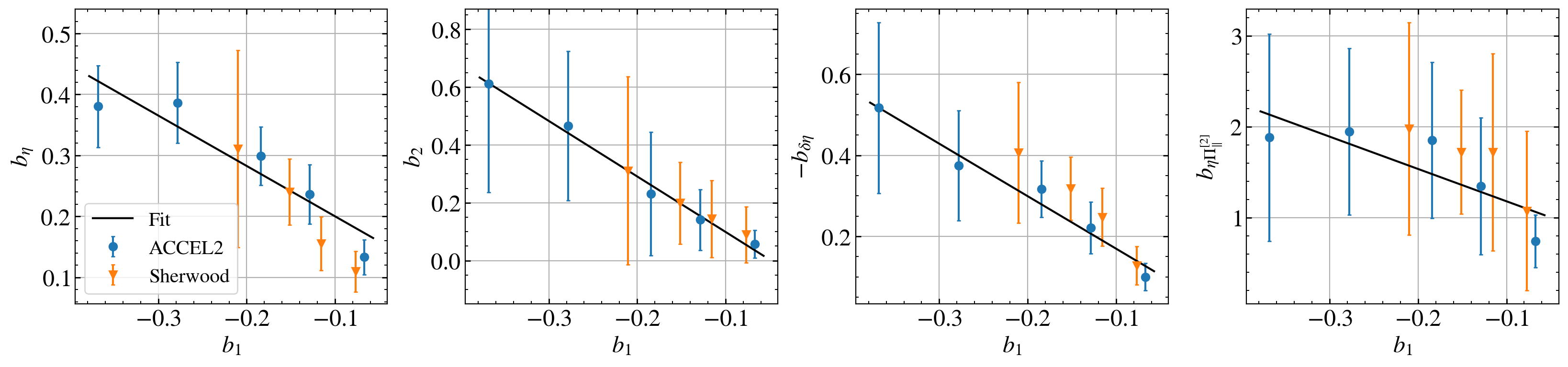} \\
    \includegraphics[width=0.75\linewidth]{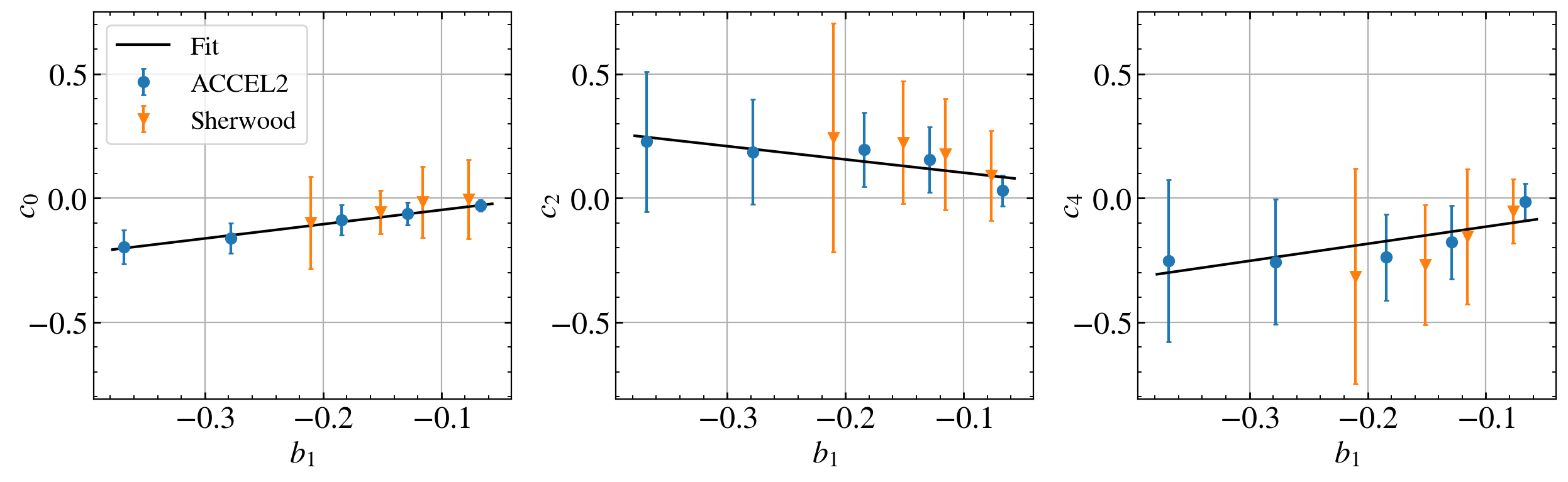}

    \caption{The bias relation with respect to $b_1$ for ({\it Top}) four EFT parameters and ({\it Bottom}) three counterterms. The same linear relation ({\it black line}) holds within the error bars between the two simulations. However, there is a minor indication of a systematic difference.
    }
    \label{fig:bias_vs_b1}
\end{figure}

The chi-squared at the best-fitting point is approximately 36 for 303 degrees of freedom across all redshift bins, indicating the 5\% noise floor overestimates the simulation uncertainties. Fig.~\ref{fig:bias_vs_z} shows four EFT bias values as a function of redshift. The redshift trend is well-described by a power law, $b_\mathcal{O}(z)=A_\mathcal{O}[(1+z)/4]^{\gamma_\mathcal{O}}$, which is mainly because the linear bias relation as postulated by ref.~\cite{ivanovEftEboss2024} holds and agrees between simulations. This $b_\mathcal{O}=b_\mathcal{O}(b_1)$ relation is shown in Fig.~\ref{fig:bias_vs_b1} for four EFT parameters and three counterterms. The biases agree between the two simulations within the error bars, with a minor indication of a systematic difference. We will address how best to parameterize these deviations later in this section. The best-fitting curve is obtained using the \accel\ data points. For $b_1$, the best-fitting values are $A_{b_1}=-0.179 \pm 0.002$ and $\gamma_{b_1}=3.25 \pm 0.08$, which can be used as priors in a real-data analysis. 

\subsubsection{Deriving stochastic terms}
As we outlined previously, $\poned$ formally requires integration of $\pthreed$ up to small scales, where the EFT is not applicable. However, we cut off the integration at the EFT scale-cut, $k_\mathrm{max}=3~$\hmpc\ such that $k^3\plin(k)/2\pi^2 \lesssim 1$ for $k<k_\mathrm{max}$ \cite{ivanovEffectiveLya2024}, and absorb the remaining UV sensitivity into stochastic terms that must be calibrated from data or simulations.

\begin{figure}
    \centering
    \includegraphics[width=0.75\linewidth]{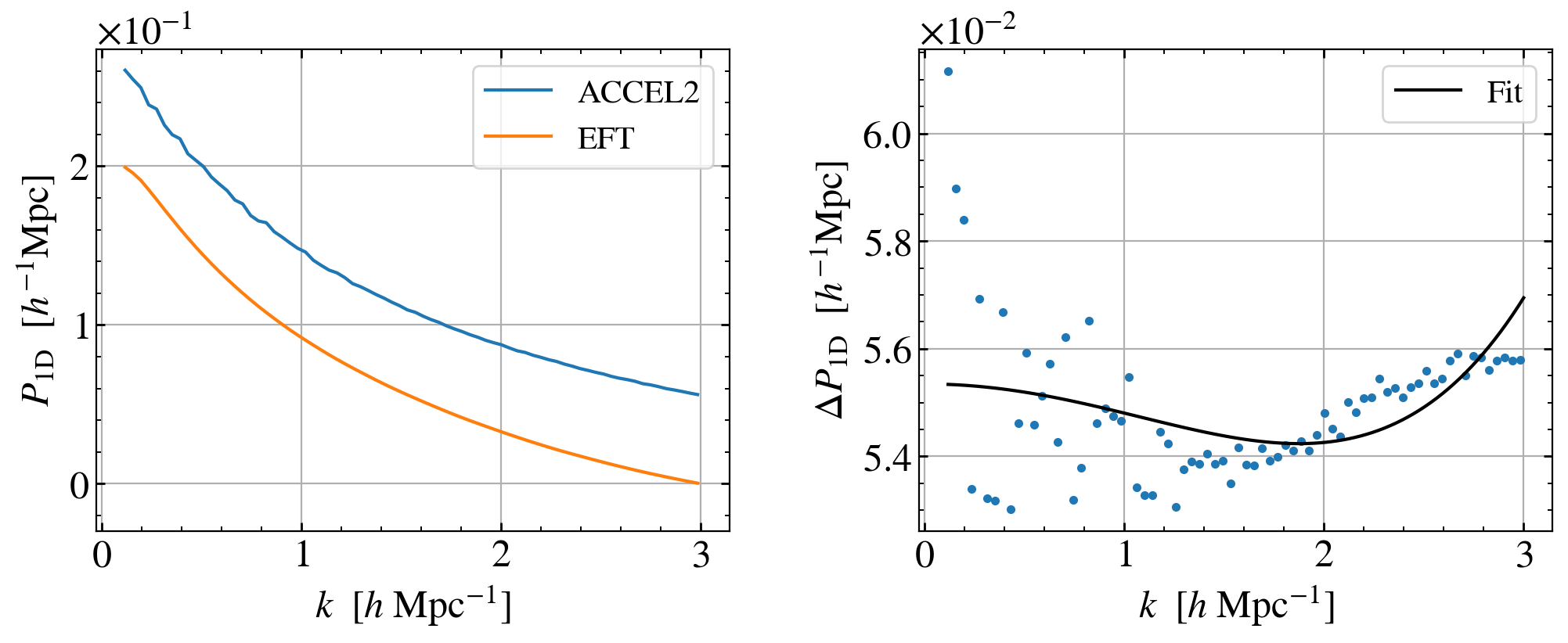}
    \caption{({\it Left}) \accel\ predictions for $\poned$ compared to the base EFT prediction at $z=2.6$. The deficit is due to the stochastic contributions. ({\it Right}) The difference between \accel\ simulations and the base EFT prediction at the same redshift. The random noise in the simulation is visible. The black line shows the best-fitting curve with three stochastic terms: $\mathcal{C}_0, \mathcal{C}_2,\text{ and } \mathcal{C}_4$. }
    \label{fig:eft_p1d_diff_stoch}
\end{figure}

We integrate the first term (i.e., $\pthreed^\mathrm{EFT}$) in Eq.~\eqref{eq:master_eft_p1d} and use the residual $\poned$ to derive the stochastic terms. The total $\poned$ measured in \accel\ simulations is compared to the EFT-only predictions in Fig.~\ref{fig:eft_p1d_diff_stoch}. 
We find that the contribution of non-stochastic EFT to the total $\poned^\mathrm{ACCEL2}$ signal at $z=2.6$ decreases monotonically with $k$: it accounts for over $70\%$ of the signal ($\poned^\mathrm{EFT}/\poned^\mathrm{ACCEL2}$) at the largest scales (lowest $k$), dropping to about $40\%$ at $k=2~$\hmpc\ and vanishing entirely ($0\%$) as expected by $k=k_\mathrm{max}=3~$\hmpc.
We fit the net power difference ($\Delta \poned$) with three stochastic terms $(\mathcal{C}_0, \mathcal{C}_2, \mathcal{C}_4)$ and find that they adequately capture the features in $\Delta \poned$. However, there is notable noise in the measured $\poned$ from simulations, especially on large scales. These are at $0.3\%$ on average and are likely correlated cosmic variance fluctuations.

\begin{figure}
    \centering
    \includegraphics[width=\linewidth]{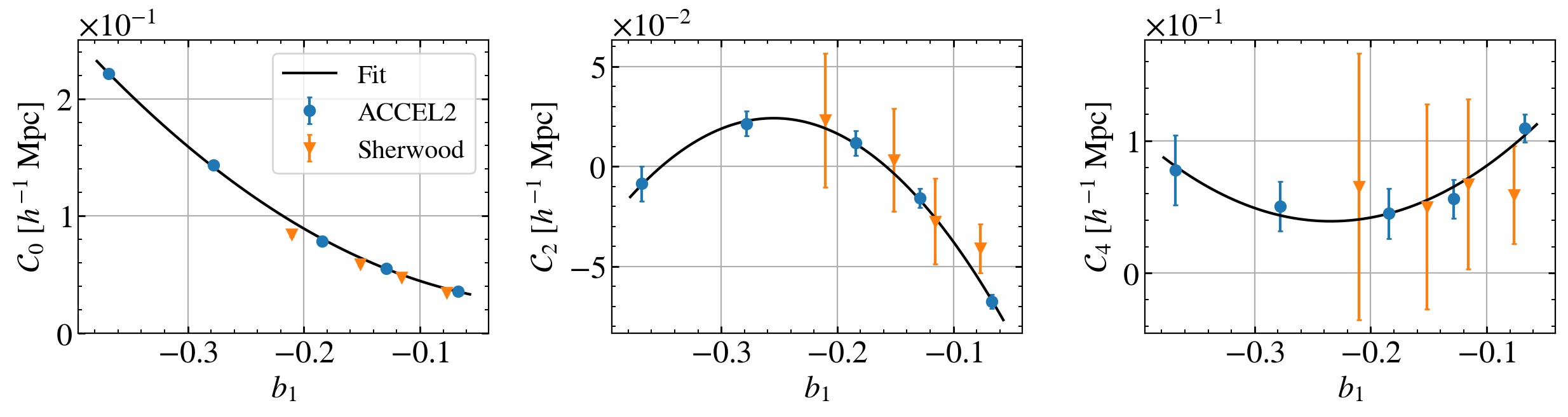}
    \caption{Stochastic counterterm measurements as a function of $b_1$ for \accel\ and Sherwood simulations. We find quadratic polynomials to be better descriptions for all $\mathcal{C}_n(b_1)$. The agreement between the two simulations is not ideal, so these parameters need to be relaxed in a real-data application.}
    \label{fig:eft_p1d_cns}
\end{figure}

The stochastic terms also obey a relation with respect to $b_1$, which can be seen in Fig.~\ref{fig:eft_p1d_cns}. We determine that all $\mathcal{C}_n(b_1)$ are better modeled with a quadratic polynomial. The agreement between simulations remains as good as for the baseline EFT parameters. We will investigate relaxing the best-fit simulation values using Gaussian priors in Section~\ref{sec:forecast}.

\subsection{Parameterizing deviations from the fiducial relation}
As evidenced by deviations between simulations and noise within the same family of simulations, the bias relations $b_\mathcal{O}=b_\mathcal{O}(b_1)$ need to be relaxed to allow a certain degree of freedom in modeling. Our EFT description so far has used 18 parameters to model $\poned$. One immediate solution is to free all these parameters with a Gaussian prior centered on the simulation mean and with a desired uncertainty. This is the most comprehensive solution in theory and readily applicable to stochastic parameters, for which we will employ the analytic template marginalization method as detailed in Section~\ref{sec:likeli}. However, given the 1D nature of our data vector, most of the bias parameters cannot be constrained even with noiseless, densely sampled data. For example, the $\beta=-b_\eta f/b_1$ parameter cannot be constrained using DESI $\poned$ observations \cite{karacayliQmleP1dDesiDr12024}, and usually cross-correlations of the \lya\ forest with quasar positions break the degeneracy of the growth rate with the velocity gradient bias \cite{cuceuCosmologyBeyondBao2021}. Additionally, we expect many degenerate directions in the bias parameter space---two different linear combinations of $\bm b_\mathcal{O}$ will move the $\poned$ in the same direction. Therefore, our proposed solution is to compress the \textit{non-stochastic} $\bm b_\mathcal{O}$ space into its most dominant orthonormal basis vectors.

In our formulation, we omit the $\mathcal{C}_n(b_1)$ terms completely in order to restrict the compression to non-stochastic parameters. The resulting basis vectors will not necessarily be orthogonal to the stochastic terms, which will likely lead to correlations. Although one could compress the full model space, motivated by the existence of empirical $\mathcal{C}_n(b_1)$ relations, the stochastic terms theoretically represent a physically distinct regime where the EFT breaks down. This is the theoretical justification behind separately compressing the base EFT space.

We investigate the base EFT space around our fiducial bias relation at fixed cosmology and per redshift bin. We also do not vary the cosmology because (1) we do not want to compress our cosmological information, and (2) we want to preserve modes that are correlated with cosmological parameters that orthonormalization might remove. We employ the Fisher matrix forecast formalism to identify the most important modes, but let us clarify our notation before proceeding with the calculations. There are two vector spaces: 1) the data vector $\poned$ with a number of measured $k$ bins ($n_k$) dimensions, and 2) the non-stochastic parameter vector space $\bm b_\mathcal{O}$ with 15 dimensions. The bold italic font is reserved for the latter in this section. Second, our goal is to find modes that produce orthonormal shifts in the data vector space. In other words, two basis vectors $\bm b_{(1)}$ and $\bm b_{(2)}$ themselves will not be orthogonal to each other $(\bm b^\mathrm{T}_{(1)} \cdot \bm b_{(2)} \neq 0)$, but their induced change in $\poned$ will be. More precisely, their corresponding element in the Fisher matrix will be zero.

The Fisher matrix is given by $\mathbf{F} = \bm{e} \cdot \mathrm{C}^{-1}\cdot \bm{e}^\mathrm{T}$, where $e_\mathcal{O}=\partial\poned/\partial b_\mathcal{O}$ are the derivatives of the model with respect to bias parameters and $\mathrm{C}$ is the covariance matrix of the data vector. Note that we ignore the quadratic contributions, which are negligible. An eigenvalue decomposition will diagonalize this matrix, and the eigenvectors with the largest eigenvalues will form the compressed orthonormal space. However, we already have a basis vector that is dictated by the parameter relations derived in the above section. So, we would like to find other vectors that induce orthogonal changes with respect to this fiducial basis vector. Let us call this fiducial vector $\bm m$, such that a coherent small step in $b_1$ corresponds to $\Delta \bm b_\mathcal{O} =\bm b_\mathcal{O}(b_1 + \mathrm{d}b_1) - \bm b_\mathcal{O}(b_1) = \bm m ~ \mathrm{d} b_1$ which results in $\mathrm{d}\poned =\bm m^\mathrm{T} \cdot \bm e ~\mathrm{d} b_1$ (note the dot product $\bm m^\mathrm{T} \cdot \bm e$ results in a vector of $n_k$ dimensions). Then, we project out this mode, $\bm v = \mathbf{F}\cdot\bm{m}$, from the Fisher matrix: $\mathbf{F}^* = \bm \Pi \mathbf{F}\bm \Pi$, where the projection matrix is defined as $\bm \Pi \equiv \left(\mathbf{I} - \bm v\bm v^\mathrm{T}/||\bm v||^2\right)$. This projection ensures that all the eigenvectors of $\mathbf{F}^*$ induce changes in $\poned$ orthogonal to our fiducial basis.\footnote{One could also project out $\bm{m}$ from $\mathbf{F}$ to satisfy two orthogonality relations.}
Finally, performing the eigenvalue decomposition on $\mathbf{F}^*$ yields eigenvectors $\bm q_n$\footnote{Note that $\bm q_n$ are not the eigenvectors of $\mathbf{F}$, but of $\mathbf{F}^*$.} and eigenvalues $\lambda_n$ which are logarithmically spaced, indicating most of the information is contained within the first few eigenvectors. To be more quantitative, dividing these eigenvalues by the largest eigenvalue $\lambda_0$, we find the second-largest eigenvalue is $\lambda_1/\lambda_0=3.4\%$, and the third-largest eigenvalue is $\lambda_2/\lambda_0=0.2\%$.

\begin{figure}
    \centering
    \includegraphics[width=\linewidth]{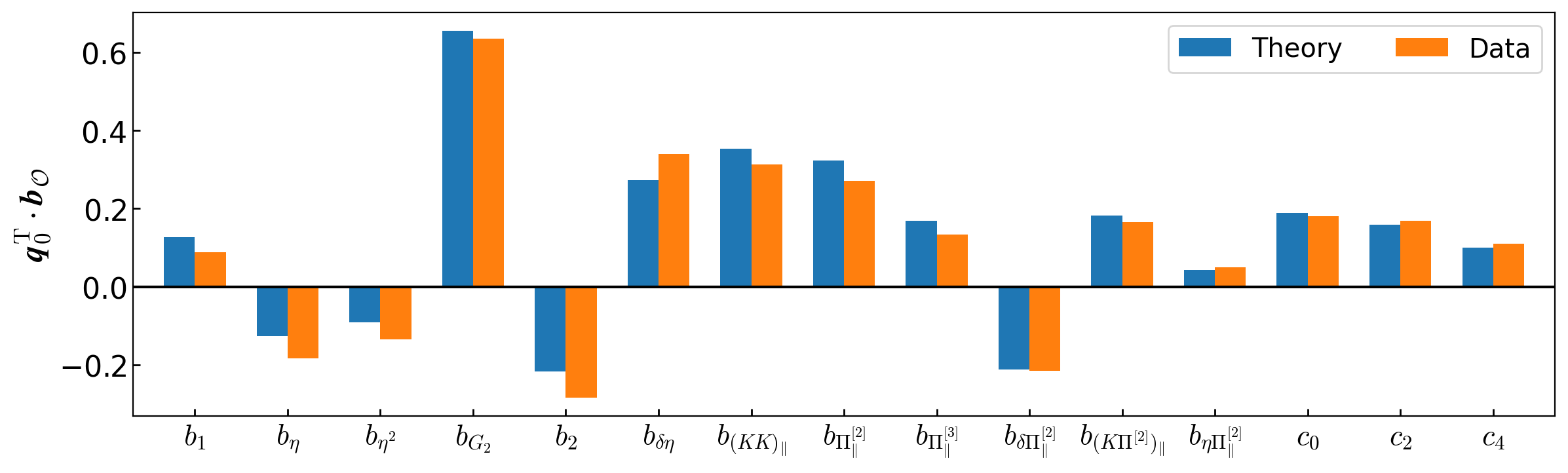}
    \caption{The component decomposition of the most dominant mode $\bm q_0$ at $z=3.0$ for a theoretical data vector in {\it blue} and for an observed data vector (set by the actual $k$ bins and covariance from the DR1 measurement) in {\it orange}. The dot product of these vectors is $99\%$, indicating a strong similarity between the two and that $\bm q_0$ is rooted in the theoretical limit.}
    \label{fig:q0_eft_theory_data}
\end{figure}
Let us start by performing this decomposition for the $z=3.0$ bin. For the data vector, we consider two options for choosing $k$ bins and the associated covariance matrix. The first possibility is to use the actual $k$ bins and covariance from the DR1 measurement to construct an ``observed" data vector. The second option we investigate is a ``theoretical" data vector that has 1000 elements equally spaced between $0.07~\text{\hmpc}<k<2.1~\text{\hmpc}$ where all $k$ bins are equally weighted with an identity matrix as its covariance. Fig.~\ref{fig:q0_eft_theory_data} shows the component decomposition of the most dominant mode $\bm q_0$. We obtain very similar vectors in both cases, with a dot product of $99\%$, indicating that $\bm q_0$ is rooted in the theoretical limit and is only slightly influenced by measurement details. This similarity worsens at some redshifts, with the smallest dot product approaching $0.96$ at $z=3.4$, which nevertheless implies a strong correlation between the two and that the theoretical limitations are the main factor determining most constrained directions.

\begin{figure}
    \centering
    \includegraphics[width=0.45\linewidth]{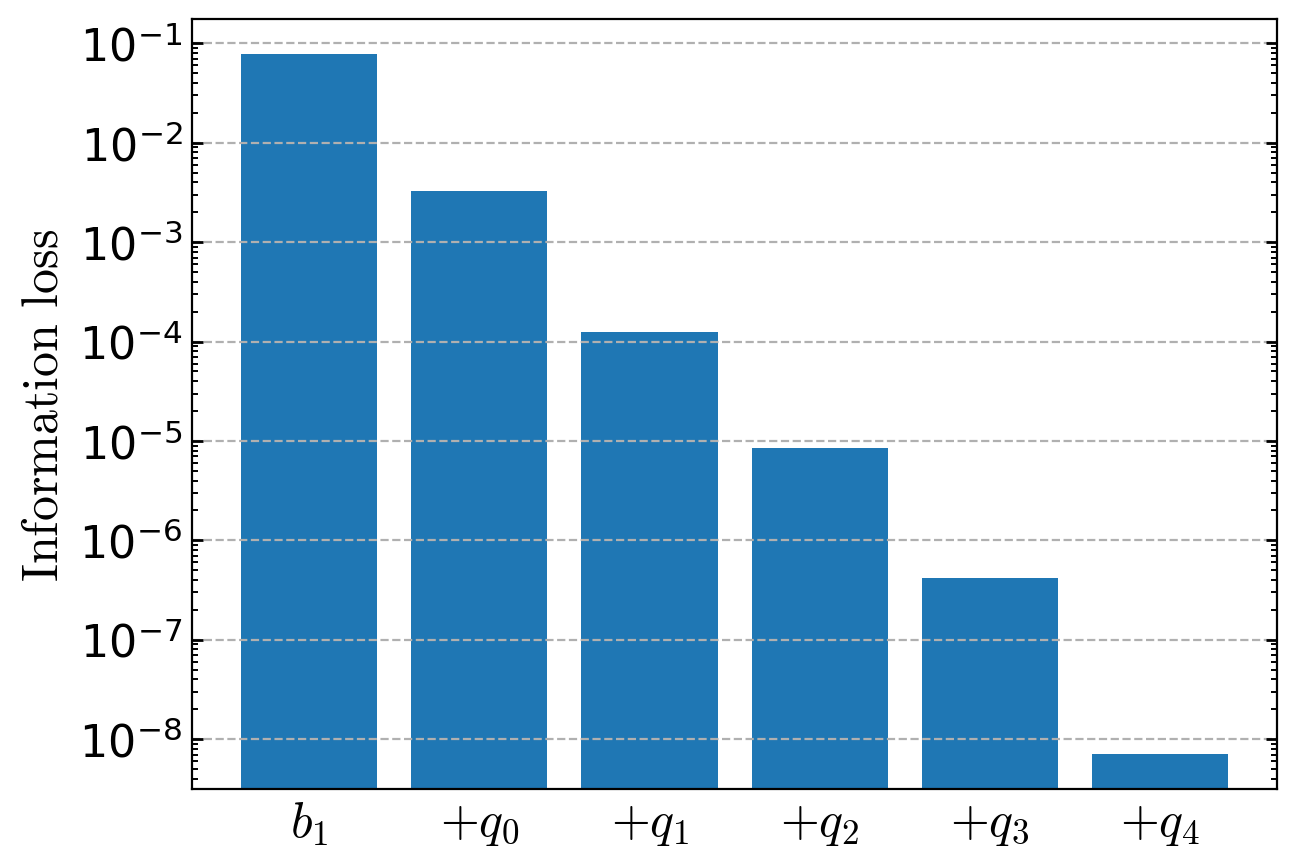} \hfill
    \includegraphics[width=0.42\linewidth]{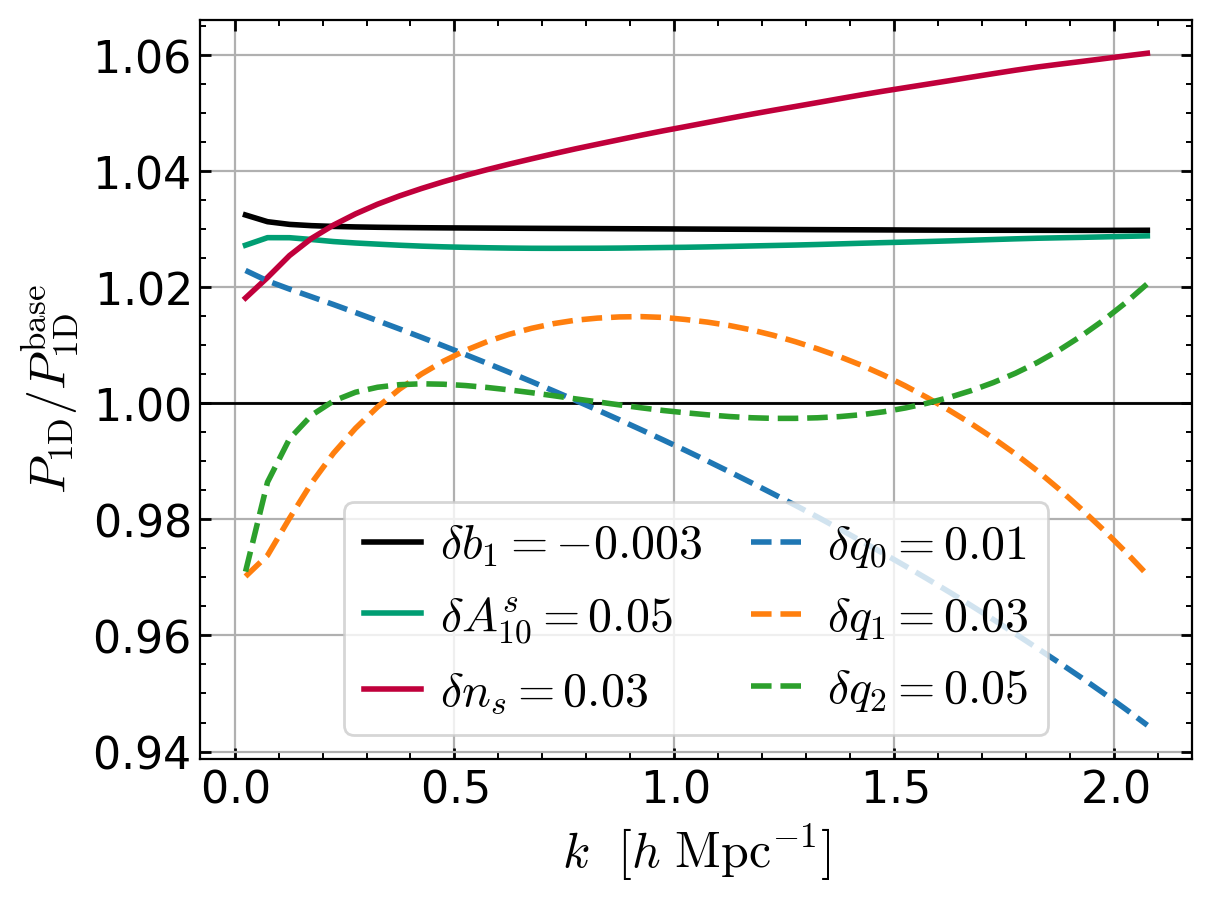}
    \caption{({\it Left}) Information loss based on the relative Frobenius norm distance between the original Fisher matrix and a reconstructed one using the compression basis at $z=3.0$. The Fisher matrix can be recovered with precision better than $10^{-5}$ using three compression vectors and better than $10^{-8}$ using five compression vectors. ({\it Right}) Response of $\poned$ in compressed directions. Nearly identical response in the fiducial $\bm b_\mathcal{O}(b_1)$ direction and $A^s_{10}\equiv \lnAs$ direction foreshadows a degeneracy between the two parameters.}
    \label{fig:infoloss_p1dresponse}
\end{figure}

To quantify the information loss due to compression, we first define the matrix $\mathbf{Q}$ where columns denote our basis vectors: $\mathbf{Q} = (\bm m, \bm q_0, \cdots,\bm q_n)$. As mentioned before, the first vector is not orthonormal to the others, and, crucially, this basis is not formed by the eigenvectors of $\mathbf{F}$. They correspond to modes $\mathbf{W} \equiv \mathbf{F}\mathbf{Q}$ in the Fisher matrix through their induced change in $\poned$. We measure the compression fidelity based on $\mathbf{W}$, after performing a singular value decomposition to column-wise orthonormalize it. We compress and reconstruct the Fisher matrix: $\mathbf{F}^\mathrm{rec} = \mathbf{W}\mathbf{W}^\mathrm{T}\mathbf{F}$, and quantify the information loss using the relative Frobenius norm distance: $||\mathbf{F}^\mathrm{rec}-\mathbf{F}||_F/||\mathbf{F}||_F$. The left panel of Fig.~\ref{fig:infoloss_p1dresponse} illustrates that the information loss exponentially decreases with the number of compression vectors at $z=3.0$ for an observed data vector, and that we can recover the Fisher matrix with $10^{-5}$ relative precision using three compression vectors.

In the right panel of Fig.~\ref{fig:infoloss_p1dresponse}, we show how the first few compression vectors change $\poned$ and compare these to the changes in two cosmological parameters: the primordial power spectrum amplitude, $A^s_{10}\equiv \lnAs$, and slope, $n_s$. Here, we calculate the new $\poned$ exactly: $\poned(\bm b_\mathcal{O}+\delta q_n \bm q_n)$ instead of relying on the finite-difference first-order derivative estimates. Compression vectors require incrementally larger $\delta q_n$ at higher orders to achieve the same order of relative change as expected. This translates to weaker constraints when fitted to data. The most concerning point is the high similarity between the fiducial $\bm b_\mathcal{O}(b_1)$ direction and $A^s_{10}$, which indicates a degeneracy between the two parameters. We will revisit the degeneracy between $b_1$ and $A^s_{10}$ and other correlations later in this section by recalculating the Fisher matrix using compressed parameters.

\begin{figure}
    \centering
    \includegraphics[width=\linewidth]{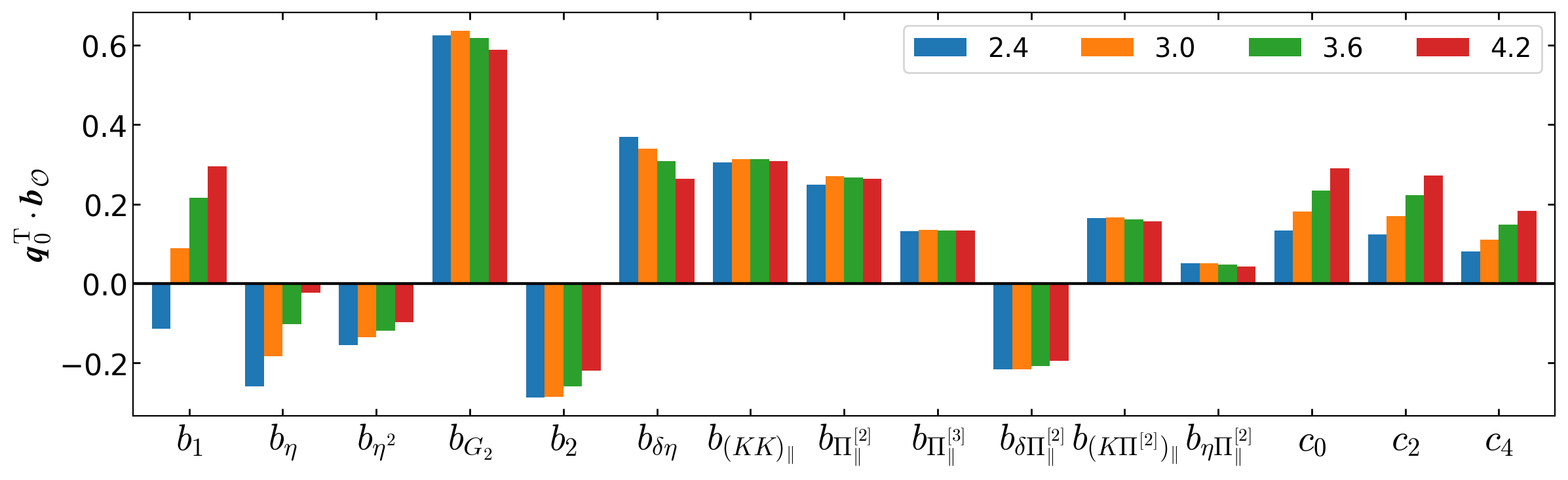}
    \caption{The component decomposition of the most dominant mode $\bm q_0$ at multiple redshift bins for an observed data vector. The correlation between $\bm q_0$ vectors gradually degrades between redshifts.
    }
    \label{fig:q0_eft_data_zbins}
\end{figure}

We now investigate the dependence of the compression vectors on redshift. In order for the compression vectors to be the most fitting for the real-data application, we derive $\bm q_n$ per redshift bin using the observed data vector. Fig.~\ref{fig:q0_eft_data_zbins} illustrates the redshift evolution of $\bm q_0$ for this case.
The correlations between $\bm q_0$ vectors gradually degrade as the distance between redshift bins increases. The lowest correlation is $72\%$ between $z=2.2$ and $z=4.4$ bins, while a more central bin, such as the $z=3.0$ bin, maintains a correlation greater than $90\%$ with all other bins. The results are similar for the other $\bm q_n$ vectors. More importantly, we find similar results for theoretical data vectors. This signifies that the main reason for these differences is being at a different point in parameter space, dictated by $\bm b_\mathcal{O}(b_1)$, rather than the use of a different covariance matrix for each redshift bin. In other words, the compression vectors can be written as a function of $b_1$: $\bm q_n = \bm q_n(b_1)$. This further indicates that orthonormality, as well as compression, may deteriorate in real-data applications when the truth diverges from our fiducial choice of $b_1(z)$. The one immediate solution to this is to 
parameterize the functional dependence of compression vectors, $\bm q_n = \bm q_n(b_1)$, which would couple the compression space and $b_1$ direction. The benefits of such an extension are not well-motivated at this stage, so we leave this to future work.

\begin{figure}
    \centering
    \includegraphics[width=\linewidth]{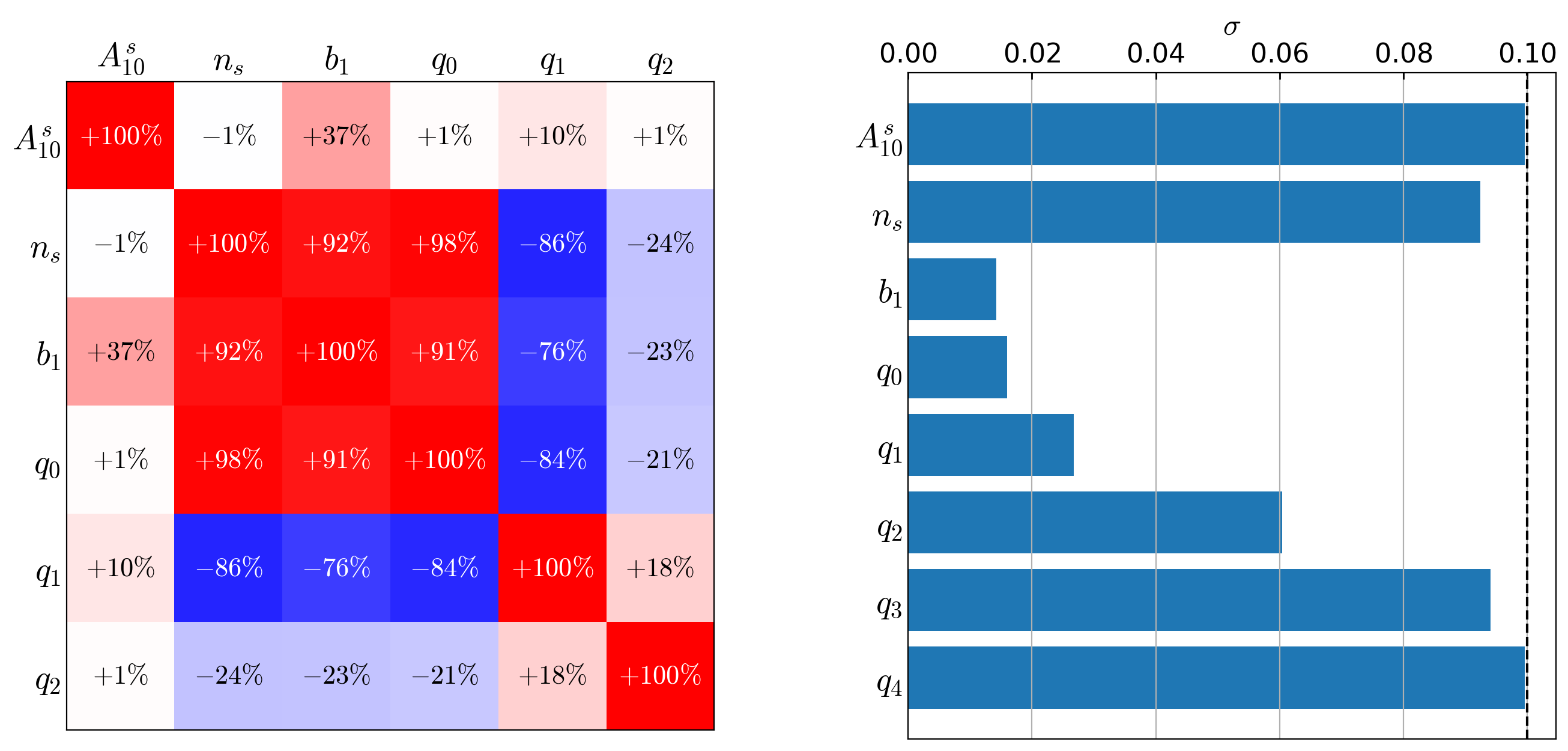}
    \caption{(Left) Estimated covariance matrix using the fiducial compression vectors $\bm q_i$ at $z=3.0$, where $A^s_{10} \equiv \lnAs$. The prior on the amplitude of the matter power spectrum decouples the otherwise completely degenerate $\bm b_\mathcal{O}(b_1)$ direction. 
    This direction and the first two compression vectors are highly correlated with $n_s$, which in turn couples otherwise uncorrelated $\bm q_0$ and $\bm q_1$ vectors. The last two vectors are not correlated with any other parameter and are therefore omitted for clarity. (Right) Estimated standard deviation for each parameter. The dashed black line marks the prior. The $A^s_{10}$, $\bm q_3$, and $\bm q_4$ cannot be constrained beyond their priors. $n_s$ performs marginally better. However, the cosmological parameters are common to all redshift bins and are expected to improve in the full analysis.}
    \label{fig:estimated_fisher_fiducialq}
\end{figure}

Let us now recalculate the Fisher matrix $(\mathbf{F}_\mathrm{zip})$ and the covariance matrix $(\mathbf{C}_\mathrm{zip})$ using compressed directions. We already alluded to the degeneracy between $b_1$ and $A^s_{10}$. This and other degeneracies destabilize the matrix inversion of $\mathbf{F}_\mathrm{zip}$, and so $\mathbf{C}_\mathrm{zip}$ estimates. To stabilize it, we assume Gaussian priors with standard deviation $\sigma_i$ for all parameters and inflate the diagonal of the Fisher matrix accordingly: $\mathbf{F}^\mathrm{zip}_{ii}\rightarrow \mathbf{F}^\mathrm{zip}_{ii} + \sigma^{-2}_{i}$, where $\sigma_i=0.1$ for all parameters. Fig.~\ref{fig:estimated_fisher_fiducialq} shows the covariance matrix at $z=3.0$ in the left panel.
$b_1$ and $A^s_{10}$ directions remain correlated, but the near-complete degeneracy is broken thanks to the prior on $A^s_{10}$. Second, the first two compression vectors become highly correlated with $n_s$, which then couples otherwise uncorrelated $\bm q_0$ and $\bm q_1$ vectors (we confirmed they are uncorrelated when cosmological parameters are fixed). Lastly, the directions of $\bm q_3$ and $\bm q_4$ are not correlated with any other parameter, so they are omitted from the figure for clarity. The standard deviation obtained from this covariance matrix is shown in the right panel. We find that $b_1$ is the most constrained parameter, followed by the compression directions in order. This demonstrates that compression order remains intact when cosmological parameters are included in the analysis. Unfortunately, this forecast also indicates that cosmological parameters cannot be constrained beyond the prior unless we use informative priors on $\bm q_0$ and $\bm q_1$. However, this is performed using a single redshift bin. In a multi-redshift-bin analysis, we expect the cosmological parameter constraints to improve as they are common to all redshift bins.

Lastly, in this section, we linearize the compressed directions in order to apply analytic template marginalization. Although these compression vectors modify biases such that $\bm b_\mathcal{O}=\bm b_\mathcal{O}(b_1) + \sum_n q_n \bm q_n$, our core, implicit, assumption so far has been $q_n\ll 1$. In other words, for large deviations, our compression scheme and orthonormality relations will break down, rendering these vectors inefficient if not faulty. One could still keep these modes in the bias parameter space using, e.g., a Gaussian prior $\mathcal{N}(0, 0.1)$ as we have done with the Fisher forecast. However, in the limit $q_n\ll 1$, a Taylor expansion of the model power spectrum will achieve similar efficiency:
\begin{equation}
    \poned\left(k; \bm b_\mathcal{O}(b_1) + \sum_n q_n \bm q_n\right) \approx \poned\left(k; \bm b_\mathcal{O}(b_1) \right) + \sum_n q_n \frac{\partial \poned}{\partial q_n}(k; \bm b_\mathcal{O}(b_1)),
    \label{eq:p1d_q_linear}
\end{equation}
with a significant advantage of analytically marginalizing over these parameters. We discuss the validity range of this approximation in Appendix~\ref{app:linear}, which turns out to be smaller than our fiducial value of $0.1$ for most of the redshift bins and for most compression directions, suggesting that stronger priors will be required to keep the linear approximation valid. Nevertheless, in order to demonstrate the power of analytic marginalization and obtain justifiable results with reasonable priors, we keep Gaussian priors of $\mathcal{N}(0, 0.1)$ for $q_n$ when quantifying the constraining power of our model in Section~\ref{sec:forecast}. Analytic marginalization is a nice benefit, but not a requirement of our framework.

\subsection{Summary of the EFT model compression\label{subsec:summary_eft}}
Let us summarize our formalism in four steps.

\begin{itemize}
    \item We first derive EFT bias parameters, $b_\mathcal{O}(z)$, at five redshift snapshots using simulations. The \accel\ simulation serves as our baseline, while the Sherwood simulation functions as an alternative. We then fit a power law for $b_1(z)$ and fit a linear function for all $b_\mathcal{O}(b_1)$ except for the stochastic terms $\mathcal{C}_0$ and $\mathcal{C}_2$. These are better described by a quadratic function.

    \item For each measurement redshift bin, we find the compression vectors, $\bm q_n$. To do that for a given redshift bin $z$, we first calculate $b_1(z)$ using the power-law redshift evolution fitted to the simulations. Then, we compute $\poned$ and its derivatives, $e_\mathcal{O}=\partial\poned/\partial b_\mathcal{O}$, using the fiducial EFT bias relations, $b_\mathcal{O}(b_1)$.

    \item This enables us to finally calculate the Fisher matrix, $\mathbf{F} = \bm{e} \cdot \mathrm{C}^{-1}\cdot \bm{e}^\mathrm{T}$, at this redshift, from which we can find the most constrained modes using an eigenvalue decomposition. Before the decomposition, we project out modes that are degenerate with the fiducial EFT relation direction. This ensures that our compression vectors induce changes orthogonal in $\poned$ to those of the $b_\mathcal{O}(b_1)$ direction.

    \item We finally linearize these compression vectors assuming $q_n\ll 1$ to enable analytic marginalization. This marginalization removes all additional, per-redshift free parameters from the sampling. The reduction in sampling space is remarkable. For example, for a $\poned$ data vector of ten redshift bins, introducing three free parameters for every bin will result in a total of 30 parameters. None of them needs to be ``free" thanks to this linearization.
\end{itemize}

What we have achieved with all these steps can be viewed as an emulator trained on simulations. This ``emulator" is analytically well-defined, but unlike its machine-learning counterparts \cite{pedersenEmulator2021, birdPriyaSimulations2023, chavesmonteroForestFlow2025}, it cannot eliminate the degeneracy between the forest bias and the amplitude of the linear matter power spectrum.\footnote{Inclusion of bispectrum \cite{zaldarriagaCorrelationsLyaForest2001, Mandelbaum:2003km} and cross-correlations with other fields (e.g., \cite{karacayli_cmb_2024}) can break this degeneracy within the EFT formalism \cite{deBelsunce:2025edy,deBelsunce:2025gci}.} However, it is readily extendable to additional degrees of freedom through higher-order compression vectors. As with any emulator, it is sensitive to the training data, which influences the fiducial bias relations, $b_\mathcal{O}(b_1)$, that is the foundation of our model.

\section{Likelihood and implementation\label{sec:likeli}}
We now describe how our formalism can be implemented in practice when multiple redshift bins are combined into a data vector, $d$. To distinguish between model-space vectors and data-space vectors, we denote model parameters in bold: $\bm \theta$, which are the EFT parameters in this work but may include other contamination parameters in general. We first start with chi-squared minimization:
\begin{equation}
    \chi^2 = (d-m(\bm \theta))^\tp \mathrm{C}^{-1} ( d - m(\bm \theta)) + (\bm \theta- \bm \mu)^\tp \mathbf{S}^{-1} (\bm \theta - \bm \mu),
\end{equation}
where we imposed Gaussian priors on model parameters centered on $\bm \mu$ with covariance matrix $\mathbf{S}$ (note that no prior on a parameter $\theta_i$ corresponds to $S_{ii}\rightarrow \infty$). In Bayesian terms, this procedure yields the maximum a posteriori (MAP) estimates of $\bm \theta$.

In the next subsections, we first provide an overview of how MAP estimates for linear templates can be analytically solved and removed from ``free" parameter space. We then outline and justify several approximations in the implementation of EFT that significantly reduce the cost of model evaluation. In the last subsection, we define a compressed cosmology following ref.~\cite{mcdonaldLinearTheoryLyaSdss2005}. Since EFT can predict $\poned$ from fundamental cosmological parameters, it provides an opportunity to test the efficiency of this postulated compression.

\subsection{Analytic template marginalization}
Our overview is similar to refs.~\cite{taylorAnalyticMethodsLikelihood2010, hadzhiyskaEfficientMarginalisation2023, sailerCrossCorrelationLRGCMB2025}. Linear model parameters can be analytically marginalized over to reduce the dimensionality of the model space. On the EFT side, the stochastic terms are ideally suited to be marginalized over using this method since they are of the form  $\mathcal{C}_n(k_\|/k_\mathrm{stoc})^n$. As discussed, we further linearized the compressed space, $\bm q_n$, representing deviations from the fiducial bias relation. Let us denote these linear parameters as $\bm \phi$, and their templates with a matrix $\mathbf{K}=\mathbf{K}(\bm \theta)$, such that $\mathbf{K}\bm\phi$ results in a vector of the same shape as the data vector $d$, where $\bm \theta$ are the remaining model parameters. For simplicity, let us define $m_0(\bm \theta)$ to be the model vector without linear terms and $r(\bm \theta) \equiv d - m_0(\bm \theta)$ as the residual vector, and consider Gaussian priors only on the linear terms. The chi-squared can be reorganized as follows:
\begin{align}
    \chi^2 &= (r(\bm \theta) - \mathbf{K}\bm\phi)^\tp \mathrm{C}^{-1} ( r(\bm \theta) -\mathbf{K}\bm\phi) + (\bm \phi- \bm \mu)^\tp \mathbf{S}^{-1} (\bm \phi - \bm \mu),\\
    &= (r^\tp \mathrm{C}^{-1} r + \bm\mu^\tp  \mathbf{S}^{-1}\bm \mu) - 2(r^\tp \mathrm{C}^{-1}\mathbf{K} + \bm \mu^\tp  \mathbf{S}^{-1}) \bm \phi + \bm \phi^\tp (\mathbf{K}^\tp \mathrm{C}^{-1} \mathbf{K} + \mathbf{S}^{-1}) \bm \phi.
\end{align}
Then, we can find $\bm \phi_\mathrm{MAP}$ that minimizes the chi-squared by calculating the zero point of its first derivative:
\begin{align}
    \left.\frac{\partial\chi^2}{\partial \bm \phi}\right|_{\bm \phi_\mathrm{MAP}} = 0 &=-2 (\mathbf{K}^\tp \mathrm{C}^{-1}r +  \mathbf{S}^{-1} \bm \mu) + 2(\mathbf{K}^\tp \mathrm{C}^{-1} \mathbf{K} + \mathbf{S}^{-1}) \bm \phi_\mathrm{MAP}, \\
    \bm \phi_\mathrm{MAP} &= (\mathbf{K}^\tp \mathrm{C}^{-1} \mathbf{K} + \mathbf{S}^{-1})^{-1} ( \mathbf{K}^\tp \mathrm{C}^{-1}r +  \mathbf{S}^{-1} \bm \mu).
\end{align}
Let us simplify the notation by defining $\bm\beta_\mathrm{MAP}\equiv \mathbf{K}^\tp \mathrm{C}^{-1}r +  \mathbf{S}^{-1} \bm \mu$ and $\mathbf{\Sigma}_\mathrm{MAP}^{-1}\equiv \mathbf{K}^\tp \mathrm{C}^{-1} \mathbf{K} + \mathbf{S}^{-1}$, such that the analytic solution for the linear parameters is $\bm \phi_\mathrm{MAP}= \mathbf{\Sigma}_\mathrm{MAP}\bm \beta_\mathrm{MAP}$. Note that a second derivative will prove $\mathbf{\Sigma}_\mathrm{MAP}$ is the analytic covariance matrix of $\bm \phi_\mathrm{MAP}$. Using this shorthand notation, we can substitute the solution back into the chi-squared expression to achieve the following compact form:
\begin{equation}
    \chi^2 = r^\tp \mathrm{C}^{-1} r + \bm\mu^\tp  \mathbf{S}^{-1}\bm \mu - \bm \beta_\mathrm{MAP}^\tp\bm \phi_\mathrm{MAP}.
\end{equation}
Typically, one sets $\bm \mu = 0$ without any loss of generality by shifting $\bm\phi \rightarrow \bm \phi + \bm \mu$ and this contribution to $m_0 \rightarrow m_0 + \bf{K}\bm{\mu}$.
Additionally, one could use the Woodbury identity \cite{woodbury1950inverting} to show this is equivalent to modifying the covariance matrix: $\chi^2 = r^\tp (\mathrm{C} + \mathbf{K} \mathbf{S} \mathbf{K}^\tp)^{-1} r$. However, we prefer computing $\bm \beta_\mathrm{MAP}$ and $\mathbf{\Sigma}_\mathrm{MAP}^{-1}$ instead of updating a large covariance matrix and its inverse at every step. This also gives us immediate access to linear parameter solutions.

The fully marginalized posterior likelihood is obtained after integration over the prior volume. This results in the following log-likelihood expression \cite{sailerCrossCorrelationLRGCMB2025}:
\begin{equation}
    \mathcal{L}_\mathrm{marg} (\bm d|\bm \theta) =-\frac{1}{2}(\chi^2 + \ln\det \mathbf{\Sigma}_\mathrm{MAP}^{-1}).
\end{equation}
If the template matrix does not depend on $\bm \theta$, then the determinant of $\mathbf{\Sigma}_\mathrm{MAP}$ is constant, in which case the maximum (marginalized posterior) likelihood and MAP estimates become equal.

\subsection{Approximations}
The exact calculation of the linear matter power spectrum and the EFT model requires substantial computational resources. We make a few well-justified approximations that significantly reduce model evaluation time.

First, we compute the linear matter power spectrum at a pivot redshift, $z_p=3.0$, for a given cosmology, and then use the linear growth function to scale it to other redshift bins. This work focuses on cases where only the primordial power spectrum parameters are varied, so we also simply rescale $\plin$ based on relative changes in these parameters instead of recomputing $\plin$.

To be flexible for cosmologies where $\Omega_m$ is also varied, we adopt the approximate linear growth function, $D(z)$, from ref.~\cite{kasaiApproximationGrowthFunction2010}, which is shown to have a small relative error of less than $0.2\%$ in the ranges $0.2\lesssim  \Omega_m\leq 1$ and $0\leq z\lesssim 1000$. The growth rate, $f(z)$, is calculated using the analytic derivative of this function.

Furthermore, fixing the growth rate to its value at $f(z_p=3.0)$ yields relative errors in $f(z)$ less than $1.5\%$ in the range $2.2 \leq z \leq 4.4$. We use this pivot growth rate to compute the EFT kernels for all redshift bins where the majority of the computational time is spent. However, note that $f$ is required both for kernel calculation and in the bias expansion (e.g., $b_\eta f\mu^2$). When calculating $\pthreed(k, \mu)$, we use the exact value of $f(z)$ since the $f(z)$ evaluation is not expensive.

\subsection{Compressed cosmology\label{sec:compress_cosmo}}
The \lya\ forest $\poned$ is measured in the redshift range $2.2\leq z \leq 4.4$, where the universe resembles an Einstein-de Sitter universe with $0.93 \lesssim \Omega_m(z) \lesssim 0.98$. The expansion history, growth function, and growth rate change at the percent level for cosmologies with $\Omega_m=0.25$ and $\Omega_m=0.35$ \cite{pedersenEmulator2021}. This motivates the proposition that $\poned$ is determined by the linear matter power spectrum at small scales, mostly by the amplitude and the logarithmic slope at a pivot scale $k_p$ \cite{mcdonaldLinearTheoryLyaSdss2005}. We will investigate the efficiency of this compression using EFT's prediction ability from fundamental cosmological parameters.

However, there has not been a consensus on this pivot point in the literature, where choices range between $k_p=0.009~$\skm\ which is the natural units for $\poned$ measurements \cite{mcdonaldLinearTheoryLyaSdss2005, desiP1dInferenceDr1_2026} to $k_p = 0.7-1.0~\text{Mpc}^{-1}$ in physical units  \cite{pedersenEmulator2021, birdPriyaSimulations2023, fernandezMultifidelityEmulator2022, fernandezCosmologicalConstraintsEboss2024, waltherEmulatingLyaP1D2024}. When the background expansion is fixed, for example, for theoretical studies such as our work, the conversion between velocity and physical distance units is a trivial constant scaling. However, when the background expansion is varied, for example, for combinations with other cosmological probes, defining the pivot scale in velocity units can prove to be more robust \cite{pedersenEmulator2021, pedersenCompressionCosmologicalLyaForest2023, desiP1dInferenceDr1_2026}.
For our work, we adopt $k_p=0.7~\text{Mpc}^{-1}$ following ref.~\cite{pedersenEmulator2021} and fit a second order polynomial to the range $k_p/2 < k <2k_p$ in logarithmic space:
\begin{equation}
    \ln P_\mathrm{lin} (k_p/2 < k <2k_p) \approx \ln \left(\frac{2\pi^2\Delta^2_p}{k_p^3}\right) + n_p \ln \frac{k}{k_p} + \frac{\alpha_p}{2}\left( \ln \frac{k}{k_p} \right)^2,
\end{equation}
where $\Delta^2_p, n_p, ~\text{and}~\alpha_p$ are the compressed cosmology parameters. To be consistent with the definition of the pivot point, we linearly interpolate $(\ln k, \ln P_\mathrm{lin})$ to 15 loglinearly-spaced points between this range such that there is an equal number of points (7) for $k_\mathrm{pivot}/2 < k$ and $k <2k_\mathrm{pivot}$. We provide the conversion between these and primordial parameters in Appendix~\ref{app:transform}.

\section{Quantifying EFT's constraining power with forecast analysis\label{sec:forecast}}

We have built some intuition for our EFT model by analyzing each redshift bin individually in Section~\ref{sec:eftmodel}. We now transfer our findings to the full data set and study the capabilities and limitations of our model when applied across multiple redshift bins. We have two major objectives in this section. (1) We found that the amplitude and the logarithmic slope of the power spectrum are degenerate with the $b_1$ parameter, indicating a challenge in constraining these cosmological parameters. We will investigate if a joint analysis of multiple time epochs could alleviate this tension. However, in cases with such strong degeneracies, a prior on the nuisance parameter is typically required to break the degeneracy in order to draw meaningful conclusions. This would be justified if the constraint on the $b_1$ parameter is substantially worse than the simulation uncertainties. Therefore, our primary objective is to understand and determine the next steps for such a scenario for the real-data application. 
(2) Our secondary objective is to quantify the constraining power of the compressed parameters and investigate their impact on cosmological parameters. An ill-constrained parameter is a likely hindrance on the path to meaningful inference. As discussed in Section~\ref{sec:eftmodel}, we investigate these parameters within their validity range, enforced by a Gaussian prior.

First, we create a data vector using the fiducial bias relations found in Section~\ref{sec:biasrelations} and impose the power-law redshift evolution for $b_1(z)=-0.179 \times [(1+z)/4]^{3.25}$. The data vector has the exact $k$ and $z$ grid as the DR1 high-SNR ($\text{SNR}>3$) measurement, with the same associated covariance matrix. Note that we do not add noise to this data vector, such that it is free from noise and modeling errors. There are ten redshift bins between $z=2.2-4.0$ in this measurement. We only use wavenumbers in the range $1.5\times 10^{-3}~\text{\skm}<k<2.0\times 10^{-2}~\text{\skm}$, which leaves 37 data points per redshift bin. The linear power spectrum is computed using \texttt{CAMB} \cite{lewisCAMB2000} with \emph{Planck} 2018 parameters \cite{collaborationPlanck2018Results2020}. We study the covariance matrix reported by the minimizer software \texttt{iminuit} \cite{jamesMinuit1975, iminuit} to quantify the cosmological constraining power.

\begin{table}
    \centering
    \begin{tabular}{|c|ccc|}
        \hline
         Parameter & $\mathcal{C}_0$ & $\mathcal{C}_2$ & $q_n$ \\
         \hline
         Prior & $\mathcal{N}(0, 0.01)$ & $\mathcal{N}(0, 0.05)$ & $\mathcal{N}(0, 0.1)$ \\
        \hline
    \end{tabular}
    \caption{Priors used in the forecast analysis. The Gaussian priors for the stochastic terms are based on integrating $\pthreed$ in linear theory, while the ones for compression directions rely on a weighted-variance propagation of Monte Carlo uncertainties from simulations.}
    \label{tab:nuissance_priors}
\end{table}

Reasonable priors for the stochastic terms can be calculated by integrating $\pthreed$ in linear theory from $k_\mathrm{max}=3~\text{\hmpc}$ to $k_\mathrm{NL}=5~\text{\hmpc}$ \cite{ivanovEffectiveLya2024}. Since the quantity of interest here is the deviation from fiducial values, we adopt 50\% of this calculation as the standard deviation of the Gaussian prior for the stochastic terms. On the other hand, our choice of priors for the compression direction is not rooted in an analytic calculation. For them, we rely on a heuristic that assumes a constant error of 0.1 for all $\bm b_\mathcal{O}$. Since compression vectors are orthonormal, this trivially projects to an error of 0.1 for all $\bm q_n$. This value is approximately justified by a weighted-variance propagation of Monte Carlo uncertainty estimates from simulations. This yields uncertainties in the range of $0.08-0.15$ for all $\bm q_n$ at $z=2.6$. In future work, the covariance matrix obtained from simulations for $\bm b_\mathcal{O}$ can be used to directly incorporate priors on the compression direction, provided it can be computed reliably. For this work, the priors used are listed in Table~\ref{tab:nuissance_priors}.

\subsection{Constraints on the amplitude and slope of the primordial power spectrum}
\begin{figure}
    \centering
    \includegraphics[width=\linewidth]{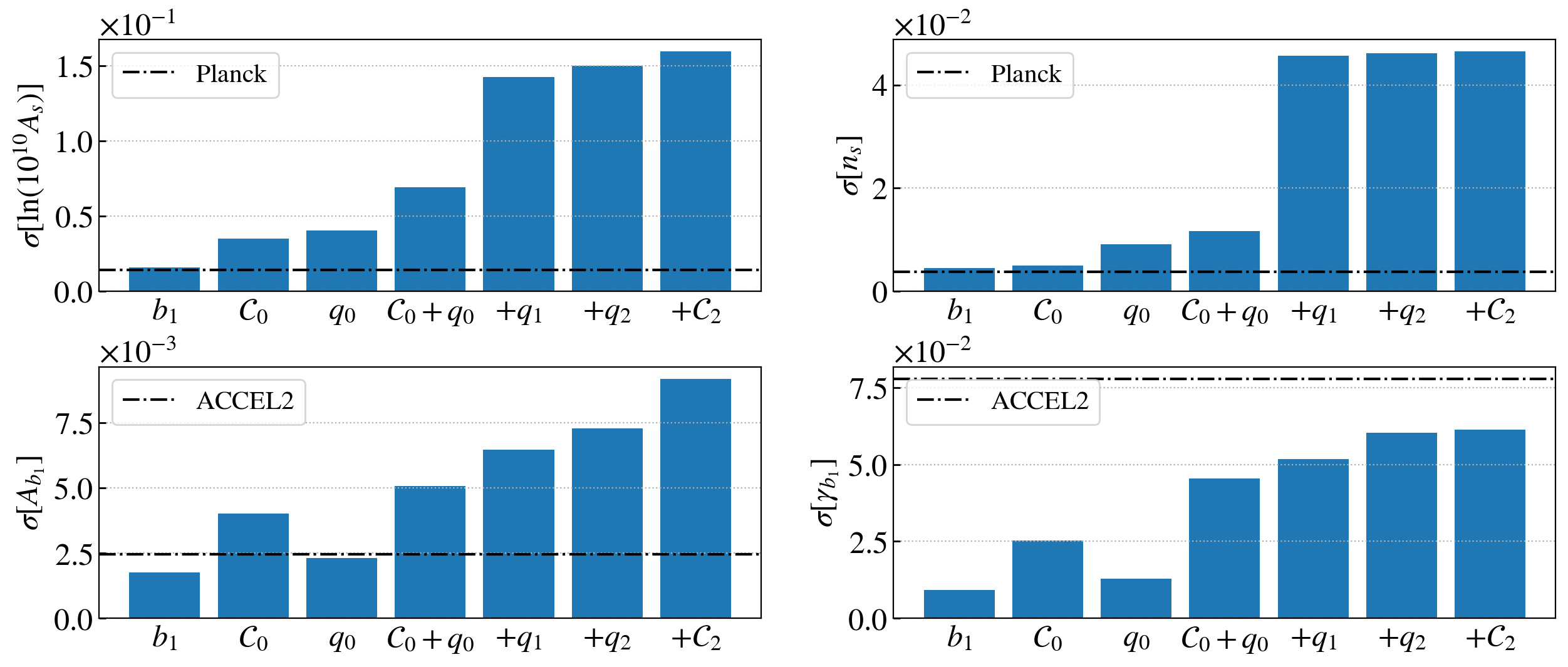}
    \caption{Estimated errors on ({\it Top}) $\lnAs$, $n_s$, and ({\it Bottom}) $A_{b_1}, \gamma_{b_1}$ from left to right, respectively, when different parameters are included in the analysis. {\it Black dotted lines} stand for the \citetalias{collaborationPlanck2018Results2020} errors for the cosmological parameters and for the error on the best-fitting curve to the \accel\ data points in Sec.~\ref{sec:biasrelations}. The leftmost column in every panel corresponds to the most constraining version of our model, where $b_1(z)$ is assumed to have a power law. As more nuisance parameters are included \emph{per redshift bin}, the errors get larger as expected. The columns that start with $+$ correspond to cumulative inclusion of parameters such that the rightmost column corresponds to the model where all nuisance parameters are included, in which case we lose about a factor of ten in constraining power.}
    \label{fig:multiz_cosmo_forecast}
\end{figure}

We start this investigation by limiting the cosmological parameters to $\lnAs$ and $n_s$ as motivated in Section~\ref{sec:compress_cosmo}. Fig.~\ref{fig:multiz_cosmo_forecast} summarizes our findings as estimated errors for $\lnAs$, $n_s$, and $A_{b_1}$ on the y-axis. The black dotted lines stand for the \citetalias{collaborationPlanck2018Results2020} errors for the cosmological parameters, and for the error on the best-fitting curve to the \accel\ data points in Section~\ref{sec:biasrelations}. 

Our baseline analysis has only four free parameters: two cosmological parameters, and two for the redshift evolution of $b_1(z)=A_{b_1} [(1+z)/4]^{\gamma_{b_1}}$. This is the first column in each panel, and represents the most constraining version of our model. Its forecast uncertainties are competitive with those of \citetalias{collaborationPlanck2018Results2020}.
We then allow deviations from the fiducial value for each redshift bin on the first stochastic term, $\mathcal{C}_0$, using a Gaussian prior $\mathcal{N}(0, 0.01)$ as justified above. This constrained freedom doubles the forecast error on the $\lnAs$ parameter, while increasing that of the $n_s$ parameter by $11\%$.
Introducing the first compression vector $q_0$ as a free variable with a Gaussian prior $\mathcal{N}(0, 0.1)$ has a larger impact on the baseline analysis results than $\mathcal{C}_0$. Having both $q_0$ and $\mathcal{C}_0$ simultaneously as free parameters increases the error on $\lnAs$ by a factor of 4.4 and the error on $n_s$ by a factor of 2.6. We then incrementally add $q_1$, $q_2$, and $\mathcal{C}_2$. All compressed parameters share the same Gaussian prior as noted previously. At every step, we lose cosmological information as expected. The amplitude constraint gets weaker consistently with every new EFT parameter introduced, whereas the most dramatic loss for $n_s$ occurs when $q_1$ is introduced. When all nuisance parameters are included in the analysis, we lose about a factor of ten in constraining power. The uncertainties for all these cases can be found in Table~\ref{tab:uncertainties}.

\begin{table}
    \centering
    \begin{tabular}{|l|c|c|}
        \hline
         & $\sigma[\lnAs]$ & $\sigma[n_s]$ \\
        \hline
        $b_1$-only & 0.016 & 0.0045 \\
        $+\mathcal{C}_0$ & 0.035 & 0.0050  \\
        $+q_0$ & 0.040 & 0.0091 \\
        $+\mathcal{C}_0+q_0$ & 0.071 & 0.0117  \\
        $\cdots+q_1$ & 0.142 & 0.0457   \\
        $\cdots+q_2$ & 0.150 & 0.0461   \\
        $\cdots+\mathcal{C}_2$ & 0.159 & 0.0465  \\
        \hline
    \end{tabular}
    \caption{The forecasted errors on $\lnAs$ and $n_s$. Our baseline analysis ($b_1$-only) has four parameters and is the most constraining version of our model. As nuisance parameters are introduced per redshift bin (i.e., ten for each new parameter), the forecasted errors increase. The bottom row corresponds to an analysis with 50 nuisance parameters added to the baseline analysis, in which case about a factor of ten is lost in constraining power.}
    \label{tab:uncertainties}
\end{table}

Given this dramatic loss in power, let us dive into our priors for the nuisance parameters. The $A_{b_1}$ parameter is constrained in a manner comparable to simulation precision within a factor of two. This can be seen in the last panel of Fig.~\ref{fig:multiz_cosmo_forecast}. It may be useful to treat the simulation precision as a prior on $A_{b_1}$ when all parameters are included. On the other hand, the $\gamma_{b_1}$ parameter is constrained beyond the simulation precision, so simulations cannot provide informative priors on this. The $q_0$, $q_1$, and $\mathcal{C}_0$ constraints remain well below their priors for all redshift bins. Based on our results, we expect $q_0$ to be well-constrained when fitted to real data. If such a fit does not strongly detect non-zero $q_0$, its prior can be strengthened to $\sim\mathcal{N}(0, 0.01)$. This is further justified by Fig.~\ref{fig:infoloss_p1dresponse}, which shows $q_0$ is a strong direction, such that $\delta q_0\sim0.01$ causing $\delta\poned\sim 0.05\times\poned$. The last two parameters, $q_2$ and $\mathcal{C}_2$, are weakly constrained, with their forecast uncertainties close to their priors. If these parameters do not improve goodness of fit or are not strongly detected, one can experiment with stronger priors or fix these parameters altogether to improve the precision of cosmological parameters in real data analysis. Conservative analyses may prefer to add compression directions incrementally until the cosmological results and their uncertainties converge. As a result of the hierarchy of compression vectors, this procedure is expected to converge unless the EFT model absorbs all of the cosmological signal into its nuisance parameters.

\subsection{Including the running of the spectral index}
We investigate how freedom in curvature affects cosmological information with a free running of the spectral index parameter, $\alpha_s$. We note that the primordial power spectrum parameters and pivot-point parameters can be mapped exactly to one another (see Appendix~\ref{app:transform}). However, for simplicity, we draw random samples of $\lnAs, n_s$, and $\alpha_s$ using a Gaussian covariance matrix and convert them to pivot-scale quantities. We will focus on two cases in this section for clarity: a $b_1$-only analysis and an all-parameter analysis. These correspond to the first and last bars in Fig.~\ref{fig:multiz_cosmo_forecast}.

\begin{figure}
    \centering
    \includegraphics[width=\linewidth]{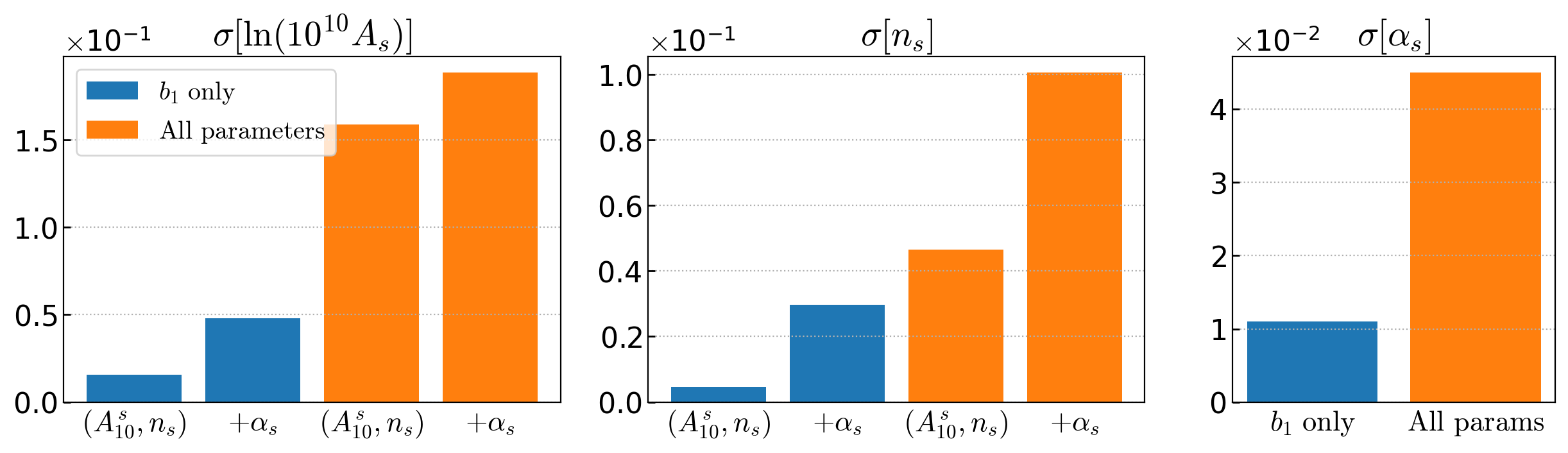}
    \caption{Estimated errors on $\lnAs$ ({\it Left}), $n_s$ ({\it Middle}), and $\alpha_s$ ({\it Right}) when the running on the spectral index, $\alpha_s$, is free. The errors increase with a free curvature, indicating that uncertainties in $\alpha_s$ propagate to those in $\lnAs$ and $n_s$. However, as shown later, this results from extrapolation beyond the pivot scale.}
    \label{fig:compressed_cosmo_alphas}
\end{figure}
Let us start with Fig.~\ref{fig:compressed_cosmo_alphas}, where we compare the errors in  $\lnAs$ and $n_s$ when $\alpha_s$ is free. The error in $\lnAs$ gets a factor of 3.1 and 1.2 larger for $b_1$-only and all-parameter analyses, respectively. The $n_s$ parameter is affected more dramatically: the error increases by a factor of 6.5 and 2.2, respectively. These indicate that (1) uncertainties in $\alpha_s$ affect uncertainties in $\lnAs$ and $n_s$, and (2) that compression directions, along with stochastic terms, weaken the constraining power to a level comparable to a free curvature.  

However, this large error inflation results from extrapolation beyond the pivot scale. When we compare the amplitude and slope at the pivot scale, we find that freedom in curvature has a small impact.
\begin{figure}
    \centering
    \begin{subfigure}[t]{.49\linewidth}
        \includegraphics[width=\linewidth]{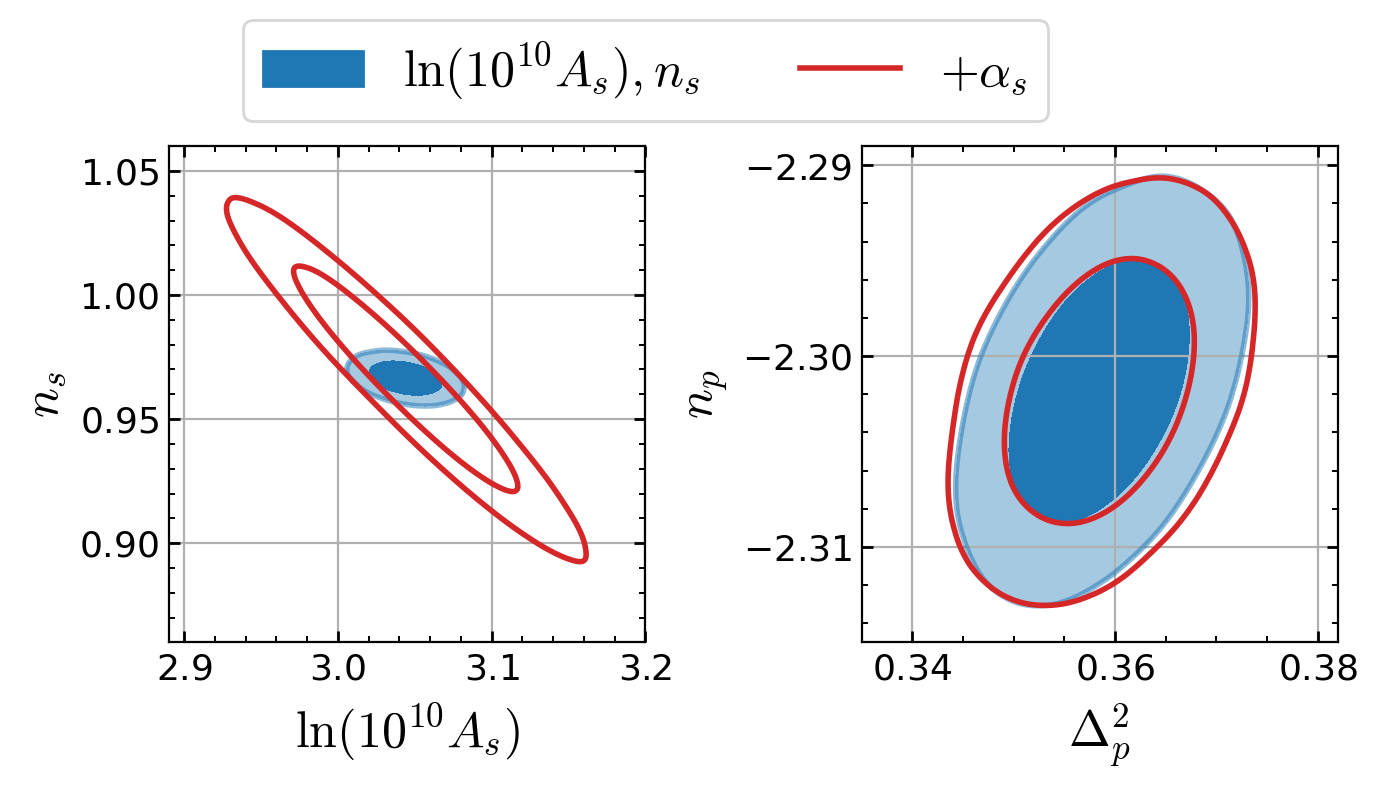}
        \caption{$b_1$-only analysis}
    \end{subfigure}
    \begin{subfigure}[t]{.49\linewidth}
        \includegraphics[width=\linewidth]{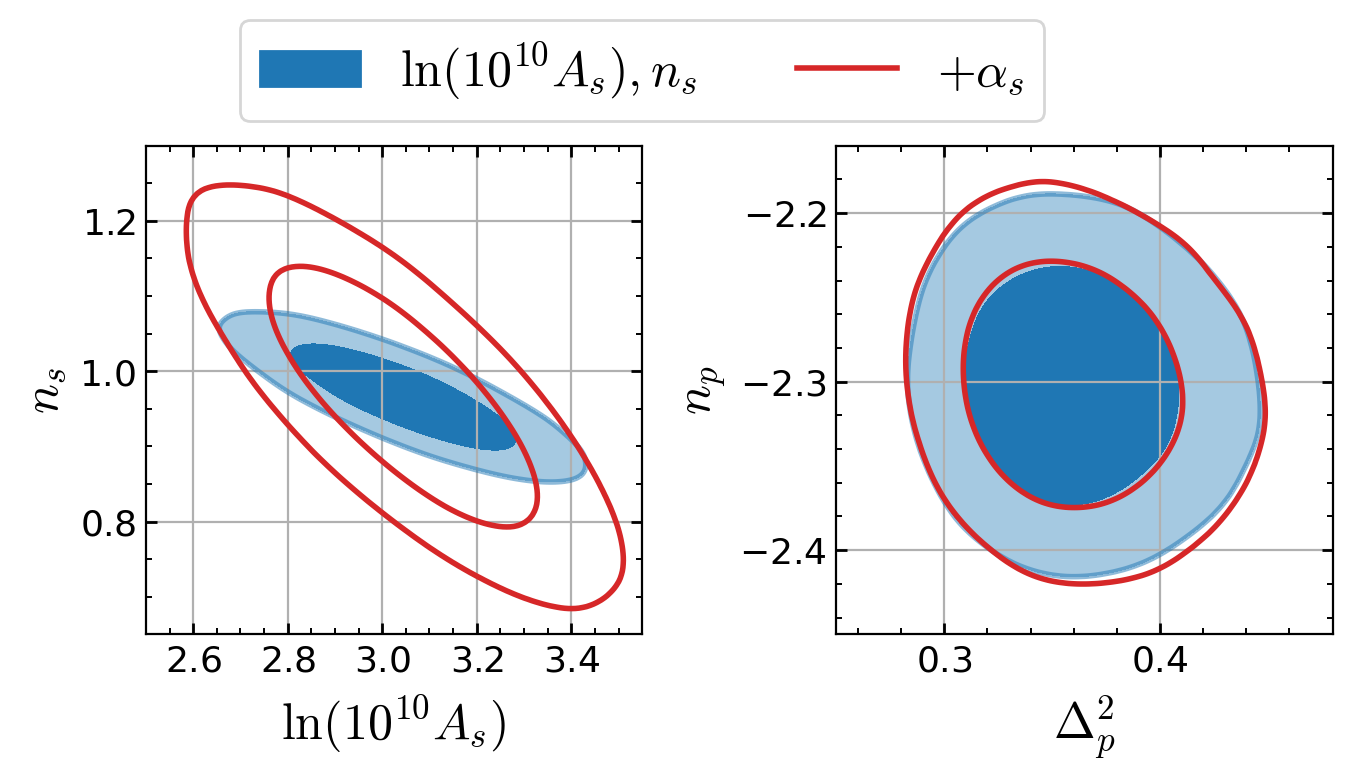}
        \caption{All-parameter analysis}
    \end{subfigure}
    \caption{Marginalized 68\% and 95\% confidence contours for primordial power spectrum and pivot-scale parameters. Blue-filled contours correspond to the baseline model with parameters $\{\lnAs,n_s\}$, while red curves show the extended model with the addition of the running of the spectral index, $+\alpha_s$.
    ({\it Left two panels}) Constraints in the $\lnAs-n_s$ plane ({\it Left}) and $\Delta^2_p-n_p$ plane ({\it Right}) for a $b_1$-only analysis. Including $\alpha_s$ significantly broadens the constraints on $\lnAs$ and $n_s$, while leaving the pivot-scale amplitude, $\Delta_p^2$, and slope, $n_p$, comparatively less affected.
     ({\it Right two panels}) Same for an all-parameter analysis. When all nuisance parameters are considered, including $\alpha_s$, the constraints on $\Delta_p^2$ and $n_p$ remain unchanged. This signifies that the compression of the \lya\ forest $\poned$ into two pivot parameters remains valid.}
    \label{fig:contour_alpha_asns_vs_dpnp}
\end{figure}
Fig.~\ref{fig:contour_alpha_asns_vs_dpnp} shows the $\lnAs-n_s$ and $\Delta^2_p-n_p$ contour plots for $b_1$-only analysis on the first two panels from the left and for all-parameter analysis on the last two panels from the right. The dramatic difference in $\lnAs-n_s$ contours with the addition of $\alpha_s$ disappears in the $\Delta^2_p-n_p$ parameterization for both cases.
When we include all nuisance parameters in the analysis, the $\Delta^2_p-n_p$ contours do not change with curvature. Therefore, for realistic data analyses, we confirm that the compression of the \lya\ forest $\poned$ into two parameters remains valid. The forecasted uncertainties are summarized in Table~\ref{tab:uncertainties_alphas}.

\begin{table}
    \centering
    \begin{tabular}{|l|c|c||c|c|c|}
        \hline
         &  \multicolumn{2}{c||}{Without curvature} & \multicolumn{3}{c|}{With curvature} \\ 
         \hline
         & $\sigma[\lnAs]$ & $\sigma[n_s]$ & $\sigma[\lnAs]$ & $\sigma[n_s]$ & $\sigma[\alpha_s]$ \\
        \hline
        $b_1$-only & 0.0156 & 0.0045 & 0.0480 & 0.0297 & 0.0110 \\
        All parameter & 0.1589 & 0.0465  & 0.1884 & 0.1005 & 0.0450 \\
        \hline \hline
        & $\sigma[\Delta^2_p]$ & $\sigma[n_p]$ & $\sigma[\Delta^2_p]$ & $\sigma[n_p]$  & $\sigma[\alpha_p]$\\
        \hline
        $b_1$-only & 0.0059 & 0.0045 & 0.0062 & 0.0046 & 0.0110 \\
        All parameter & 0.0331 & 0.0464  & 0.0337 & 0.0484 & 0.0450 \\
        \hline
    \end{tabular}
    \caption{The forecasted errors on $\lnAs-n_s$, and $\Delta^2_p-n_p$ without curvature and with curvature (free running on the spectral index, $\alpha_s$). The forecasted uncertainties for the pivot parameters remain largely unaffected.}
    \label{tab:uncertainties_alphas}
\end{table}

Our final forecasted precision for $\Delta^2_p$ is $10\%$ and for $n_p$ is $2.0\%$ when all nuisance parameters and the curvature are included in the analysis. The precision in $\Delta^2_p$ is marginally worse than the precision of the official DESI DR1 $\poned$ inference results of $8.7\%$ presented in ref.~\cite{desiP1dInferenceDr1_2026}. On the other hand, our forecasted precision in $n_p$ is a factor of $2.5$ worse than the official precision of $0.8\%$.
To its advantage, the DESI DR1 $\poned$ inference analysis uses a Gaussian process emulator that directly connects cosmology to $\poned$ by training on simulations: $\poned=\poned(\bm\theta_\mathrm{cosmo})$. It also includes numerous systematics that we have not considered. When these systematics, such as high-column-density systems, are marginalized over, we may lose more constraining power. We leave the study of these to future work, where we apply our formalism to real data.

\section{Discussion\label{sec:discuss}}

In this work, we have proposed a model compression for the EFT power spectrum computation for the 1D \lya\ flux power spectrum. The key motivation for our work is to reduce the space of sampled EFT parameters by effectively removing their degenerate combinations from the model. To that end, we have used the Fisher-matrix compression approach, allowing us to identify the principal components of EFT templates. Using the EFT parameterization around the \accel\ simulation-based correlations of EFT parameters and the linear bias coefficient $b_1$, we found that without the assumptions about the redshift-dependence of EFT parameters, the bulk of the EFT model parameter space is captured by only three combinations of EFT parameters that appear in three dimensions.

Specifically, we have produced a projection of the cosmological sensitivity of the EFT-based \lya\ analysis of the DESI DR1 data and found that even when all the EFT parameters are varied per each redshift bin, we can constrain the amplitude ($\Delta^2_p$) and the logarithmic slope ($n_p$) of power spectrum of linear mass fluctuations at the pivot scale ($k_p=0.7~\text{Mpc}^{-1}$) to $10\%$ and $2.0\%$, respectively, which is only slightly worse than the emulator-based analysis of DESI DR1 \cite{desiP1dInferenceDr1_2026}.  These emulators learn $\poned=\poned(\bm\theta_\mathrm{cosmo})$ from simulations, and using perturbation-theory terminology, implicitly exploit additional information encoded in $b_1(\bm\theta_\mathrm{cosmo})$ and break the strong degeneracy between $b_1$ and $(\Delta^2_p, n_p)$. In this regard, EFT is a more conservative approach since the bias parameters' connection to cosmology is not learned and is instead freely fit to the data.
Our constraints, however, are saturated when using only 6 EFT parameters per redshift bin  (the linear bias $b_1$, stochastic counterterms $\mathcal{C}_{0,2}$, and three principal components of non-linear EFT parameters $q_{0,1,2}$). This opens the possibility of significantly boosting the efficiency of EFT-based full-shape analyses of $\poned$, while retaining substantial flexibility in the EFT model. Our analysis thus demonstrates the substantial constraining power of EFT, even under a conservative treatment of its parameters.

We note that our approach is based solely on compressing the theory space. In principle, one can consider simultaneously suppressing the theory and model spaces along the lines of MOPED~\cite{Heavens:1999am} or SVD-based compression \cite{philcoxFewerMocksLessNoise2021}. In addition, our constraints can be improved by making additional assumptions, e.g., about the smooth time-dependence of EFT coefficients, or by using stronger priors on the EFT principal components from simulations. Similarly, our approach can also be applied to inference using the three-dimensional power spectrum. We leave the exploration of these options, as well as the application of our technique to DESI data, for future work. 

\paragraph{Software.}
We use the following commonly-used software in \texttt{python} analysis: \texttt{astropy}\footnote{\url{https://www.astropy.org}}
a community-developed core \texttt{python} package for Astronomy \citep{astropy:2013, astropy:2018, astropy:2022},
\texttt{numpy}\footnote{\url{https://numpy.org}}
an open source project aiming to enable numerical computing with \texttt{python} \citep{numpy},
\texttt{scipy}\footnote{\url{https://scipy.org}} an open-source project with algorithms for scientific computing,
\texttt{numba}\footnote{\url{https://numba.pydata.org}}
an open source just-in-time (JIT) compiler that translates a subset of \texttt{python} and \texttt{numpy} code into fast machine code,
Finally, we make plots using
\texttt{matplotlib}\footnote{\url{https://matplotlib.org}}
a comprehensive library for creating static, animated, and interactive visualizations in \texttt{python}
\citep{matplotlib}.

\acknowledgments
We thank Andreu Font-Ribera for helpful comments and for providing the Sherwood transmission files.

NGK acknowledges support from the United States Department of Energy, Office of High Energy Physics under Award Number DE-SC0011726. This work is supported by the National Science Foundation under Cooperative Agreement PHY-2019786 (The NSF AI Institute for Artificial Intelligence and Fundamental Interactions, \url{http://iaifi.org/}).

\appendix

\section{EFT kernels}\label{app:EFT_kern}
The one-loop Lyman-$\alpha$
power spectra are given by
\be 
\begin{split}
& P_{22}=\int 2\frac{d^3q}{(2\pi)^3}[K_2(\k-\q,\q)]^2P_{\rm lin}(|\k-\q|)P_{\rm lin}(q)~\,,\\
& P_{13} =3K_1(\k)P_{\rm lin}(k) \int \frac{d^3q}{(2\pi)^3}K_3(\k,-\q,\q)P_{\rm lin}(q)\,,
\end{split}
\ee 
where we have used:
\be
\label{eq:K2full}
\begin{split}
& K_1(\k) = b_1-b_\eta f\mu^2\,,\\
& K_2(\k_1,\k_2)=\frac{b_2}{2}+b_{\mathcal{G}_2}\left(\frac{(\k_1\cdot \k_2)^2}{k_1^2 k_2^2}-1\right)+b_1F_2(\k_1,\k_2)  -b_\eta f\mu^2 G_2(\k_1,\k_2) - fb_{\delta \eta}\frac{\mu_2^2+\mu_1^2}{2} +b_{\eta^2}f^2\mu_1^2\mu_2^2\\
& +b_1f\frac{\mu_1\mu_2}{2}\left(\frac{k_2}{k_1} + \frac{k_1}{k_2}\right)
-b_\eta f^2\frac{\mu_1\mu_2}{2}\left(\frac{k_2}{k_1}\mu_2^2 + \frac{k_1}{k_2}\mu_1^2\right) + b_{(KK)_\parallel}\left(\mu_1\mu_2 \frac{(\k_1\cdot \k_2)}{k_1k_2}
-\frac{\mu_1^2+\mu_2^2}{3}+\frac{1}{9}
\right)\\
& + b_{\Pi^{(2)}_\parallel}\left(\mu_1\mu_2 \frac{(\k_1\cdot \k_2)}{k_1k_2}+\frac{5}{7}\mu^2 \left(1-\frac{(\k_1\cdot \k_2)^2}{k_1^2 k_2^2}\right)\right)\,,
\end{split} 
\ee
with $\mu_i=(\hat{\bf z}\cdot \hat{\k}_i)$, $\hat {\bf z}$ denoting 
the line-of-sight direction unit vector,  
$f=d\ln D_+/d\ln a$ ($D_+$ is the growth factor),
and we have used the
usual density and velocity kernels from standard cosmological perturbation theory:
\be
\begin{split}
F_2(\k_1,\k_2) &= \frac{5}{7} + \frac{2}{7}\frac{(\k_1\cdot \k_2)^2}{k_1^2k_2^2} + \frac{1}{2}\frac{\k_1\cdot \k_2}{k_1k_2}\bigg(\frac{k_1}{k_2} + \frac{k_2}{k_1}\bigg)\,\,, \\
G_2(\k_1,\k_2) &= \frac{3}{7} + \frac{4}{7}\frac{(\k_1\cdot \k_2)^2}{k_1^2k_2^2} + \frac{1}{2}\frac{\k_1\cdot \k_2}{k_1k_2}\bigg(\frac{k_1}{k_2} + \frac{k_2}{k_1}\bigg) \,\,,
\end{split}
\ee
The general expression for the $K_3$ kernel is quite cumbersome. 
Below we present it only for the 
kinematic configurations that appear in the one-loop power spectrum integrals: 
\be 
\label{eq:K3full}
\begin{split}
& \int_{\q} K_3(\k,\q,-\q) 
P_{\rm lin}(q)
\\
=\,&
b_1
\int_{\q} F_3(\k,\q,-\q)P_{\text{lin}}(q)
-
f b_\eta\mu^2
\int_{\q} G_3(\q,-\q,\k)P_{\text{lin}}(q)
\:+ 
\int_{\q}
\left[
1 - \left(\khat\cdot\qhat\right)^2
\right]
P_{\rm lin}(q)
\\
&\:\times
\Bigg\{
\frac{4}{21}
(
5b_{\mathcal{G}_2}
+ 
2b_{\Gamma_3})
\left[\left(\frac{(\k-\q)\cdot\q}{|\k-\q|q}\right)^2 - 1\right]
- 
\frac{2}{21} f b_{\delta\eta}
\left[
   \frac{3(k_\parallel-q_\parallel)^2}{|\k-\q|^2}
   +
   \frac{5q_\parallel^2}{q^2}
\right]
\\
&\quad\:
+\frac47 f^2 b_{\eta^2}
\frac{q_\parallel^2}{q^2}
\frac{(k_\parallel-q_\parallel)^2}{|\k-\q|^2}
+
\frac{20}{21} b_{(KK)_\parallel}
\left[
    \frac{(\k\cdot\q-q^2)(k_\parallel-q_\parallel)q_\parallel}{|\k-\q|^2q^2} 
    - \frac13\frac{(k_\parallel-q_\parallel)^2}{|\k-\q|^2} - \frac13 \frac{q_\parallel^2}{q^2} + \frac19
\right]
\\
&\quad\:+\frac{10}{21}
b_{\Pi_\parallel^{[2]}}
\frac{(\k\cdot\q-q^2)}{|\k-\q|^2}
\frac{(k_\parallel-q_\parallel)^2}{q^2}
+
\frac{10}{21}
\left[
b_{\delta\Pi_\parallel^{[2]}} 
-\frac13 b_{(K\Pi^{[2]})_\parallel}
- 
f b_{\eta\Pi_\parallel^{[2]}} 
\frac{q_\parallel^2}{q^2}
\right]
\frac{(k_\parallel-q_\parallel)^2}{|\k-\q|^2}
\\
&\quad\:+
\frac{10}{21}
b_{(K\Pi^{[2]})_\parallel}
\frac{(\q\cdot\k-q^2)}{q|\k-\q|}
\frac{q_\parallel(k_\parallel-q_\parallel)}{q|\k-\q|}
+
\frac{10}{21}
f b_{\Pi^{[2]}_\parallel}
\frac{q_\parallel(k_\parallel-q_\parallel)^3}{q^2|\k-\q|^2}
\\
&\quad\:
+
(b_{\Pi_\parallel^{[3]}}+2b_{\Pi_\parallel^{[2]}})
\Bigg[
\frac{13}{21}
\frac{\k\cdot\q-q^2}{|\k-\q|^2}
\frac{q_\parallel(k_\parallel-q_\parallel)}{q^2} 
-
\frac{5\mu^2}{9}
\left[\left(\frac{(\k-\q)\cdot\q}{|\k-\q|q}\right)^2 - \frac{1}{3}\right]
\Bigg]
\\
&\quad\:+
\frac{2}{21}f b_1 
\left[
5
\frac{q_\parallel(k_\parallel-q_\parallel)}{q^2}
+
3
\frac{q_\parallel(k_\parallel-q_\parallel)}{|\k-\q|^2}
\right]
-
\frac27 
f^2 
b_\eta
\frac{q_\parallel(k_\parallel-q_\parallel)}{q^2|\k-\q|^2}
\left[
(k_\parallel-q_\parallel)^2
+
q_\parallel^2
\right]
\Bigg\}\,.
\end{split}
\ee
The expressions for the standard perturbation theory kernels $F_3$ and $G_3$ can be found e.g. in Ref.~\cite{Bernardeau:2001CPT}.

\begin{figure}
    \centering
    \includegraphics[width=0.45\linewidth]{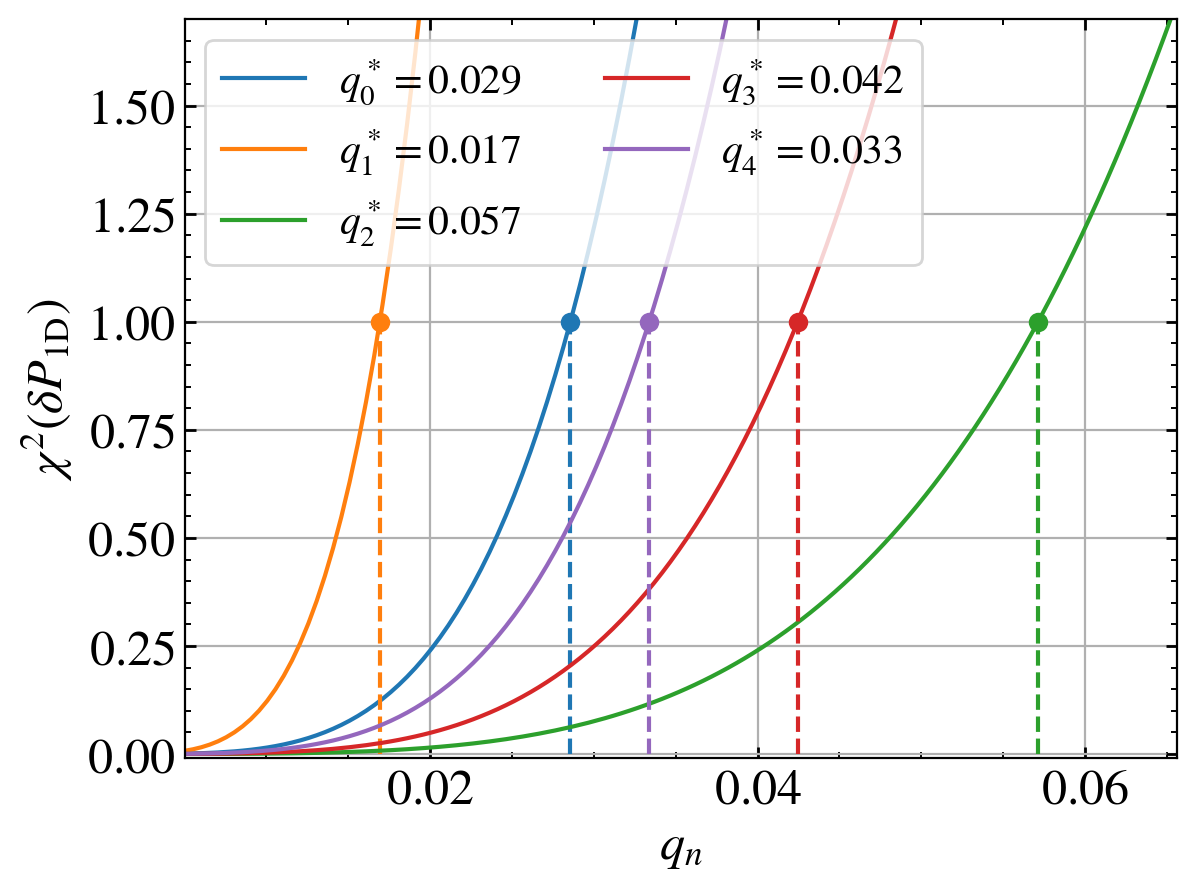}
    \includegraphics[width=0.45\linewidth]{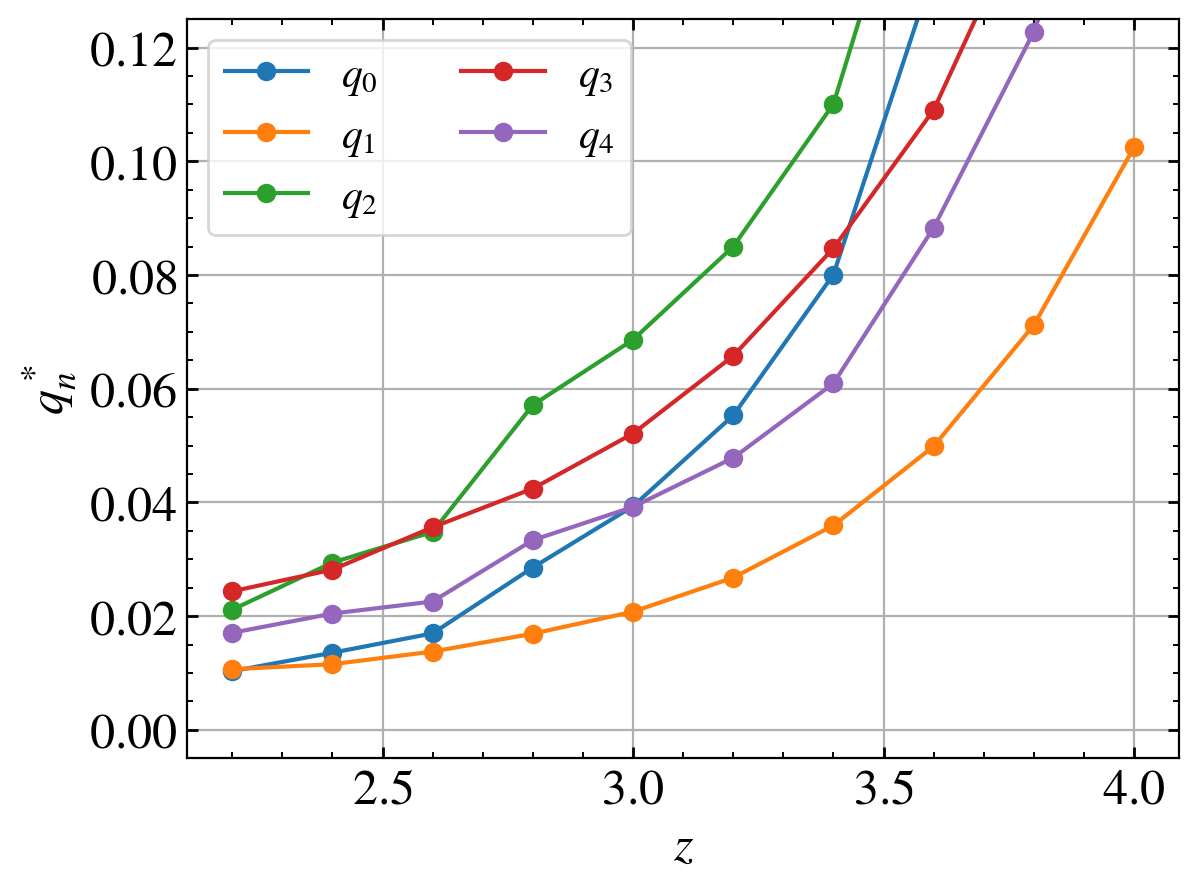}
    \caption{({\it Left}) $\chi^2$ profiles for each compression direction at $z=2.8$. The strengths of compression directions do not correlate with their non-linearity. ({\it Right}) The one-sigma boundary, $q_n^*$, where $\chi^2(q_n^*)=1$, as a function of redshift. The boundary expands as the measurement error gets larger at higher redshifts.}
    \label{fig:chi2_profile_linear}
\end{figure}

\section{Validity range of the linear approximation\label{app:linear}}
The Taylor expansion in Eq.~\eqref{eq:p1d_q_linear} ties non-linear bias parameters of EFT into linear compression directions. This approximation breaks down eventually. To assess the validity range of linearization, we quantify the error between the exact $\poned$ calculation,
\begin{equation}
    \poned^\mathrm{exact}=\poned\left(k; \bm b_\mathcal{O}(b_1(z)) + \sum_n q_n \bm q_n\right),
\end{equation}
and the linear approximation,
\begin{equation}
    \poned^\mathrm{lin}=\poned\left(k; \bm b_\mathcal{O}(b_1(z))\right) + \sum_n q_n \frac{\partial \poned}{\partial q_n}(k; \bm b_\mathcal{O}(b_1(z))),
\end{equation}
for each compression direction at a redshift bin $z$. The relevant tolerance scale is set by the measurement error, such that we adopt $\chi^2=\delta \poned^\mathrm{T}\mathrm{C}^{-1} \delta \poned$ as a metric and define $\chi^2<1$ (i.e., the one-sigma region) as the validity range of the linear approximation. The 1D boundary, $q_n^*$, is defined as $\chi^2(q_n^*)=1$.

Fig.~\ref{fig:chi2_profile_linear} shows $\chi^2$ profiles for each compression direction at $z=2.8$ in the left panel. It turns out that the strength of the compression directions and their non-linearity do not correlate. For example, the figure reveals that the $q_1$ direction reaches the one-sigma boundary at a lower value than the $q_0$ direction: $q_1^* < q_0^*$, and also $q_2^*$ is the largest of all five directions. As the measurement error gets larger at higher redshifts, the validity range expands proportionally, as shown in the right panel of Fig.~\ref{fig:chi2_profile_linear}. Overall, the one-sigma validity region emerges to be smaller than the $\mathcal{N}(0, 0.1)$ priors for most of the redshift bins and for most compression directions.

\section{Transformation between primordial to pivot parameters\label{app:transform}}
The changes in primordial parameters around a fiducial cosmology (i.e., $\delta\bm\theta_s\in \delta A_s, \delta n_s, \delta \alpha_s$) can be transformed to changes in pivot parameters (i.e., $\delta\bm\theta_p=\delta \Delta^2_p, \delta n_p, \delta \alpha_p$). Equating the induced changes in $\plin$ with respect to deviations in these parameter spaces results in the following:
\begin{align}
    \ln \frac{P(k; \bm\theta_s + \delta \bm\theta_s)}{P(k; \bm\theta_s)} &= \ln \frac{P(k; \bm\theta_p + \delta \bm\theta_p)}{P(k; \bm\theta_p)} \\
    \delta \ln A_s + \delta n_s \ln \frac{k}{k_s} + \frac{\delta \alpha_s}{2} \left(\ln \frac{k}{k_s}\right)^2 &= \delta \ln \left(\frac{2\pi^2\Delta^2_p}{k_p^3}\right) + \delta n_p \ln \frac{k}{k_p} + \frac{\delta \alpha_p}{2}\left( \ln \frac{k}{k_p} \right)^2.
\end{align}
We substitute $\ln \frac{k}{k_s}=\ln \frac{k}{k_p}\frac{k_p}{k_s}$ to the left hand side and define $\Gamma \equiv \ln \frac{k_p}{k_s}$.
\begin{align}
    \delta \ln A_s + \delta n_s \left(\Gamma + \ln \frac{k}{k_p}\right) + \frac{\delta \alpha_s}{2} \left(\Gamma + \ln \frac{k}{k_p}\right)^2 &= \cdots \\
    \left(\delta \ln A_s + \Gamma \delta n_s + \Gamma^2 \frac{\delta \alpha_s}{2}\right) + \left( \delta n_s + \Gamma \delta \alpha_s \right) \ln \frac{k}{k_p} + \frac{\delta \alpha_s}{2} \left(\ln \frac{k}{k_p}\right)^2 &= \cdots,
\end{align}
which forms a simple system of linear equations. The forward and backward transforms are as follows:
\begin{align}
    \begin{pmatrix}
        \delta \ln \Delta^2_p \\
        \delta n_p \\
        \delta \alpha_p
    \end{pmatrix} &= \begin{pmatrix}
        1 & \Gamma & \Gamma^2/2 \\
        0 & 1 & \Gamma \\
        0 & 0 & 1
    \end{pmatrix} \begin{pmatrix}
        \delta \ln A_s \\
        \delta n_s \\
        \delta \alpha_s
    \end{pmatrix}, \\
    \begin{pmatrix}
        \delta \ln A_s \\
        \delta n_s \\
        \delta \alpha_s
    \end{pmatrix} &= \begin{pmatrix}
        1 & -\Gamma & \Gamma^2/2 \\
        0 & 1 & -\Gamma \\
        0 & 0 & 1
    \end{pmatrix} \begin{pmatrix}
        \delta \ln \Delta^2_p \\
        \delta n_p \\
        \delta \alpha_p
    \end{pmatrix}.
\end{align}
Note that the natural logarithm term in the amplitude makes the transformation formally non-linear. However, for small changes in the amplitude, the transformation is approximately linear.

\bibliographystyle{JHEP}
\bibliography{references,more_refs}

\end{document}